\documentclass[aip,jcp,reprint]{revtex4-2}

\usepackage{amsmath}
\usepackage{graphicx}
\usepackage{dcolumn}
\usepackage{bm}
\usepackage{mathrsfs}
\usepackage{color}
\usepackage{natbib}

\begin{document}

\title{Nonequilibrium statistical thermodynamics of a small quantum system coupled to quantum reservoirs}
\author{Jeremy Schofield}%
\email{deceased}
\author{Raymond Kapral}%
\email{r.kapral@utoronto.ca}
\affiliation{Chemical Physics Theory Group, Department of Chemistry, University of Toronto, Toronto, Ontario, Canada, M5S 3H6}

\date{\today}

\begin{abstract}
A statistical mechanical formulation where nonequilibrium averages of selected quantum observables are given exactly as averages over local density operators is used to describe the dynamics and thermodynamics of a small quantum system coupled to reservoirs. The local density operators are functionals of constraint variables that play the role of thermodynamic forces that satisfy a set of equations of motion. A focus of the work is on the evolution of the average internal energy of the quantum subsystem that may be subject to external autonomous driving that does work on the system, and on entropy production in the system. The nonequilibrium statistical thermodynamics the quantum subsystem coupled to reservoirs, as well as that for the isolated subsystem and reservoirs, are discussed. When the subsystem is coupled to large quantum reservoirs, the effects of the quantum subsystem on the reservoirs can be captured through boundary conditions on the reservoir equations, and these boundary conditions are derived. The formulation is illustrated by considering the heat transfer through a system comprising a quantum system coupled to two heat reservoirs with different temperatures.
\end{abstract}

\date{\today}
\maketitle

\section{Introduction}

The extension of thermodynamic concepts for open quantum systems is being actively pursued, and various approaches to describe how the first and second laws of thermodynamics can be extended to such systems are described in literature devoted to this topic.~\cite{gemmer2009quantum,vandenbroeck2013stochastic,binder2018thermodynamics,deffner2019quantum,Alicki2019,Talkner2020Colloquium,strasberg2022quantum,kurizki2022thermodynamics,Gaspard2022,Limmer2024,Bittner2026-GeometricThermo,Potts2026Quantum}

Both the quest to understand how thermodynamics can be extended to microscopic scales and experimental studies of molecular-level quantum devices have provided stimulus for research in this area. The derivations of classical and quantum fluctuation formulas have provided deep insight into the microscopic basis of the thermodynamic laws and their generalizations.~\cite{Evans1993,Jarzynski1997,Crooks2000,Andrieux2004,Sevick2008,Esposito2009,Seifert_2012,Esposito2012Stochastic,Ding2022UnifiedThermo}
Also, applications related to single molecule force experiments~\cite{Ritort2007Nonequilibrium} and quantum computation~\cite{Haack2026thermodynamic} have prompted research on their statistical and thermodynamic properties.

Entropy plays a central role in the extensions of the laws of thermodynamics to small quantum systems, and a number of suggestions for its definition have been given in the literature.~\cite{Swendsen_2015entropy} The difficulties in defining entropy stem from the observation that familiar definitions such as the von Neumann or information entropy are invariant under the unitary evolution of quantum mechanics and cannot correspond to the thermodynamic entropy. As a consequence other formulations that do not suffer from this difficulty have been used for this purpose and their properties investigated.~\cite{entropy-esposito-2010,Safranek2019entropy,Safranek2019entropy2,Seifert2020Entropy,Strasberg_2021micro-entropy,Elouard2026quantumthermo}
In many instances formulations are based on Jaynes maximum entropy construction~\cite{Jaynes1957,Jaynes1957_II}, where the system's probability distribution is obtained by maximizing the information entropy subject to constraints that fix average values of observables. Effective time-dependent temperatures are often a part of such formulations.~\cite{cugliandolo2011effective,Strasberg2021Clausius}

In this work we study the dynamics and derive generalized thermodynamic relations for a quantum system coupled to nonequilibrium reservoirs using a statistical mechanical formulation~\cite{M56,M58,mclennan1963formal,R66,R67,P68,Robertson1970,OL79} where a local density operator is constructed by maximizing an entropy functional subject to constraints that yield the exact average values of a selected set of observables. In this formulation, which is in the spirit of Jaynes' maximum entropy principle, the equations of motion for the constraint functions, which play the role of thermodynamic forces conjugate to the fluxes of the observables, are constructed with the aid of time-dependent projection operators. Using this formalism, we construct the exact evolution equation for the internal energy of a quantum subsystem, subject to external autonomous driving and in contact with nonequilibrium quantum reservoirs, and the entropy production in the entire system. Expressions for the work and heat are obtained and analyzed. For the isolated quantum system in the absence of reservoirs, the work is related to the change in internal energy and computed for irreversible and reversible processes for a specific work protocol. For a small quantum system coupled to large reservoirs, it is shown how the reservoir energy fields may be obtained by using boundary conditions at the interface between the subsystem and reservoir that are derived by decomposing the reservoir dynamics into coupling and bulk domains. The results are illustrated by computing the heat flux through a system where quantum subsystem lies between two quantum reservoirs at different temperatures.


\section{Quantum subsystem and reservoirs}
We consider the dynamics and thermodynamics of a small quantum subsystem in contact with quantum reservoirs that maintain it under nonequilibrium conditions. The reservoirs are assumed to be large compared to the system and interact with it at its boundaries through short-range intermolecular forces. In addition, the quantum subsystem may be subjected to external time-dependent driving that does work on the system. The Hamiltonian operator for the entire quantum system is
\begin{equation}\label{eq:Hamiltonian}
\widehat{\mathsf{H}}(\lambda_t)=\widehat{H}_S(\lambda_t)+\widehat{V}_c +\widehat{H}_R \equiv \widehat{H}_S(\lambda_t) +\widehat{H}_0,
\end{equation}
which is a sum of the subsystem Hamiltonian, $\widehat{H}_S(\lambda_t)$, the coupling operator $\widehat{V}_c$
and reservoir Hamiltonian $\widehat{H}_R$. While the subsystem Hamiltonian can have various forms, here we choose $\widehat{H}_S(\lambda_t)= \widehat{H}_S +\lambda_t \widehat{B}$, where $\widehat{H}_S$ is the Hamiltonian of the subsystem in the absence of driving, while the external driving term $\lambda_t \widehat{B}$ is the product of the operator $\widehat{B}$ and $\lambda_t$, an autonomous time-dependent function that determines the work protocol. Also, it is convenient for some purposes to consider the coupling term along with the reservoir Hamiltonian and define $\widehat{H}_0=\widehat{H}_R +\widehat{V}_c$ as in Eq.~(\ref{eq:Hamiltonian}). For some applications the reservoir may comprise $n$ independent reservoirs with corresponding coupling operators, in which case we have $\widehat{H}_R =(\widehat{H}_{R1}, \widehat{H}_{R2}. \dots, \widehat{H}_{Rn})$ and $\widehat{V}_c=(\widehat{V}_{c1},\widehat{V}_{c2},\dots, \widehat{V}_{cn})$.

The Liouville equation for the density operator of the entire system is
\begin{equation}
\partial_t \widehat{\rho}(t)= -\frac{i}{\hbar} [\widehat{\mathsf{H}}(\lambda_t), \widehat{\rho}(t)] \equiv -i\widehat{\mathcal{L}}(\lambda_t) \widehat{\rho}(t),
\end{equation}
and its formal solution can be written in terms of the time-ordered exponential operator,
\begin{equation}
\widehat{\mathcal{U}}^\dagger(t,0)=\mathcal{T}_+ \exp\Big(-\int_0^t d{t_1} \; i\widehat{\mathcal{L}}(\lambda_{t_1}) \Big),
\end{equation}
with smaller time arguments to the right of larger ones, and is given by
\begin{equation}
\widehat{\rho}(t)= \widehat{\mathcal{U}}^\dagger(t,0) \widehat{\rho}(0).
\end{equation}

Rather than attempting to solve the Liouville equation for the full density operator, we instead focus on the nonequilibrium expectation values $\bm{a}(\bm{r},t)$ of a selected set of observables $\{{\widehat{\bm{A}}(\bm{r})}\}$, whose elements may depend on the spatial position $\bm{r}$ within the system. These averages are defined by
\begin{equation}\label{eq:exact-av}
\bm{a}(\bm{r},t) = {\rm Tr} \big[\widehat{\bm{A}}(\bm{r}) \widehat{\rho}(t)\big],
\end{equation}
where ${\rm Tr}$ denotes the trace over the quantum degrees of freedom. ( Throughout the following discussion,
nonequilibrium expectation values are denoted by lowercase symbols.) The precise form of the density operator $\widehat{\rho}(t)$ and the trace operation depend on the physical setting.

Although the equations of motion for the nonequilibrium averages $\bm{a}(\bm{r},t)$ can be derived for any choice of operators comprising the set $\{\widehat{\bm{A}}(\bm{r}) \}$, for this investigation it is necessary that the set includes the subsystem observables whose dynamics are of primary interest, as well as those  reservoir operators whose densities capture reservoir features that are important for its interaction with the subsystem.

In particular, for large reservoirs the coarse-grained densities of conserved reservoir operators evolve slowly in spatial regions far from the quantum subsystem because their time derivatives are proportional to spatial gradients, which are small on macroscopic scales. It follows that the set of variables should include local densities, such as reservoir particle number, momentum, and energy densities. More generally, the set may also contain slow non-conserved variables, such as those that are found in chemically reacting systems, and in systems with spontaneously broken continuous symmetries or in the vicinity of thermodynamic critical points.

We are interested in situations where the small quantum subsystem in the absence of coupling to reservoirs or external driving is described globally and evolves by unitary dynamics governed by the Hamiltonian $\widehat{H}_S$. The formalism can be generalized to encompass both a more detailed description of quantum dynamics as well as large quantum subsystems~\cite{large-sys,Haack2017}.

Based on these considerations it is convenient to break up the set of operators into two sets, $\{\widehat{\bm{A}}(\bm{r}) \} = \{ \widehat{\bm{A}}_s, \widehat{\bm{A}}_0 (\bm{r}) \}$, where the first set of operators $\widehat{\bm{A}}_s$ are global operators of the quantum subsystem, and the second set of operators $\widehat{\bm{A}}_0 (\bm{r})$ are density operators that depend on the reservoir degrees of freedom and the spatial position in the system.

The Hamiltonian of the quantum subsystem and its dependence external driving are essential observables, and we choose $\widehat{\bm{A}}_s=(\widehat{B}, \widehat{H}_S)$. For the systems considered here, the most important reservoir operator is its local Hamiltonian, $\widehat{H}_R(\bm{r})$. In particular, we choose $\widehat{A}_0(\bm{r})=\widehat{H}_0(\bm{r})$, since it is useful to include the coupling between the reservoir and the subsystem in the local Hamiltonian operator, $\widehat{H}_0(\bm{r})=\widehat{H}_R(\bm{r})+\widehat{V}_c(\bm{r})$. This set of variables $\widehat{\bm{A}}(\bm{r})=( \widehat{\bm{A}}_s, \widehat{A}_0(\bm{r}) )$ is appropriate for the description of a small quantum subsystem where a global description of its properties is of most interest, while the choice of the single operator $\widehat{H}_0(\bm{r})$ is appropriate for a class of reservoirs that do not involve either fluid flows or particle fluxes. As noted earlier, a wider class of open quantum systems can be studied using this formalism if additional subsystem and reservoir operators are included in the operator set.

\section{Nonequilibrium average values of operators}
The exact average values of any set of operators $\{\widehat{\bm{A}}(\bm{r})\}$ in Eq.~(\ref{eq:exact-av}) can be expressed as an average over a local density operator for an open quantum system defined by
\begin{equation}\label{eq:local-equil}
\hat{\rho}_L(t)= e^{\widehat{\bm{A}}(\bm{r}) \ast \bm{\phi}(\bm{r},t)}/\mathcal{Z}(t),
\end{equation}
where $\mathcal{Z}(t)={\rm Tr}\big[  e^{\widehat{\bm{A}}(\bm{r}) \ast \bm{\phi}(\bm{r},t)}\big]$ is the partition function that guarantees its normalization. The conjugate $\bm{\phi}$ fields are constructed so that the exact nonequilibrium averages of the $\widehat{\bm{A}}(\bm{r})$ operators are equal to their local nonequilibrium averages:~\cite{M56,M58,mclennan1963formal,R66,R67,P68,Robertson1970,OL79,MG2023,QC-like}
\begin{align}\label{eq:a-def}
\bm{a}(\bm{r},t) &={\rm Tr} \big(\widehat{\bm{A}}(\bm{r}) \hat{\rho}(t) \big)  \\
&= {\rm Tr} \big(\widehat{\bm{A}}(\bm{r}) \hat{\rho}_L(t)\big)\equiv \langle \widehat{\bm{A}}(\bm{r})\rangle_t , \nonumber
\end{align}
where $\langle \cdots \rangle_t$ denotes an average over $\hat{\rho}_L(t)$. In Eq.~(\ref{eq:local-equil}) and below, the notation $\hat{\bm{A}} (\bm{r}) \ast \phi(\bm{r} ,t)$ is a short-hand representation for
\begin{equation}
\hat{\bm{A}} (\bm{r}) \ast \phi(\bm{r} ,t) = \hat{\bm{A}}_s \cdot \bm{\phi}_s (t) + \int d\bm{r} \, \hat{A}_0(\bm{r},t)  {\phi}_0(\bm{r},t),
\end{equation}
where $\bm{\phi}_s (t)$ and $\bm{\phi}_0(\bm{r},t)$ are Lagrange multipliers for the subsystem and reservoir density operators, respectively.  We will frequently suppress the spatial and time dependence and write $\hat{\bm{A}} (\bm{r}) \ast \phi(\bm{r} ,t) = \hat{\bm{A}} \ast \bm{\phi}$ to simplify the notation.

In later sections of this paper where we deal with equations for the specific variables and operators in these sets, it is convenient to write the constraint fields in a form that reflects their physical character, so we let
\begin{equation}
\bm{\phi}_s(t)=-(\alpha(t),\beta_S(t)), \;\phi_0(\bm{r},t)=-\beta_0(\bm{r},t),
 \end{equation}
 where $\beta_S(t)$ and $\beta_0(\bm{r},t)$ are generalized inverse temperature variables and $\alpha(t)$ is associated with the driving operator $\widehat{B}$. The average fields of the operators are denoted by \begin{equation}
 \bm{a}_s(t)=(b(t),e_S(t)), \; a_0(\bm{r},t)=e_0(\bm{r},t).
 \end{equation}
In this notation we can write
\begin{equation}
\widehat{\bm{A}}(\bm{r}) \ast \bm{\phi}(\bm{r},t)= - \widehat{B} \alpha(t) -\widehat{H}_S \beta_S(t)  - \widehat{H}_0(\bm{r}) \ast \beta_0(\bm{r},t),
\end{equation}
where we again use the $\ast$ symbol for the spatial integral over the product of local fields, $\widehat{H}_0(\bm{r}) \ast \beta_0(\bm{r},t)= \int d\bm{r} \;\widehat{H}_0(\bm{r}) \beta_0(\bm{r},t)$.

The local density operator can be derived using a generalization of the standard Gibbs approach to obtain the equilibrium density~\cite{Gibbs,Jaynes1957,Jaynes1957_II}. In particular, the form of the local density operator in Eq.~(\ref{eq:local-equil}) can be obtained by maximizing an entropy functional $S(t)$,
\begin{align} \label{eq:entropy}
S(t) = -k_B {\rm Tr} \left[  \rho_L(t) \ln \rho_L(t)  \right],
\end{align}
with respect to the functional form of $\rho_L(t)$ subject to a set of the locally imposed constraints~\cite{OL79} defined in Eq.~(\ref{eq:a-def}).

The $\bm{\phi}(\bm{r},t)$ fields conjugate to the dynamical variables $\widehat{\bm{A}}(\bm{r})$ correspond to time-dependent and possibly spatially-dependent Lagrange multipliers that enforce the constraint conditions.  As in equilibrium, the Lagrange multipliers and the local nonequilibrium averages of the dynamical variables are  related through functional derivatives of the partition function ${\cal Z}(t)$ and the entropy functional via,
\begin{align}
\bm{a}(\bm{r},t) = \frac{\delta \ln {{\cal{Z}} (t)}}{\delta \bm{\phi}  (\bm{r},t)}, \; \bm{\phi}(\bm{r},t) = -\frac{1}{k_B} \frac{\delta S(t)}{\delta \bm{a}(\bm{r},t)}
\end{align}

Although more general forms of the local density operator can be constructed by imposing constraints on the nonequilibrium averages of multilinear densities as well~\cite{SO94}, we shall not consider such extensions here. Their primary effect is to generate mode-coupling corrections to the transport coefficients appearing in the evolution equations for averages of linear densities, and these corrections are typically subleading.

\subsection{Evolution equations}\label{subsec:density-evolution}
We now turn to the derivation of the equations of motion for the average fields $\bm{a}(\bm{r},t)$ and their conjugate thermodynamic fields $\bm{\phi}(\bm{r},t)$, whose solutions determine the nonequilibrium state of the system.~\cite{other-operators} We present the main results below, while further details of the derivations are given in Appendices~\ref{app:projectors} and \ref{app:av-value-eqs}.

The evolution equation for the average observables can be obtained by taking the time derivative of the first equality in Eq.~(\ref{eq:a-def}). If the set of operators $\{ \widehat{\bm{A}}(\bm{r}) \}$ is independent of the driving protocol $\lambda_t$, as is the case for our set of chosen operators, we obtain:
\begin{equation}
\label{eq:dadt}
\dot{\bm{a}}(\bm{r},t)= {\rm Tr} \big( \widehat{\bm{A}}(\bm{r}) \partial_t\widehat{\rho}(t)\ \big)={\rm Tr} \big( \widehat{\bm{\mathcal{J}}}_{A}(\bm{r}))\widehat{\rho}(t) \big).
\end{equation}
In the second equality we moved the time derivative from the density to the operator and defined its flux by $\widehat{\bm{\mathcal{{J}}}}_{A}(\bm{r}) =  i\widehat{\mathcal{L}}(\lambda_t) \widehat{\bm{A}}(\bm{r})=\frac{i}{\hbar} [\widehat{\mathsf{H}}(\lambda_t), \widehat{\bm{A}}(\bm{r})]$. To complete the derivation we require a relation between the full and local density operators.

This relation can be obtained with the aid of a time-dependent projector~\cite{R66,OL79},
\begin{eqnarray}\label{eq:Pdag-proj}
\widehat{\mathcal{P}}^\dagger(t)\widehat{f}(t)
&=& {\rm Tr}  \big(\widehat{f}(t)\big)\widehat{\rho}_L(t)\\
&&+{\rm Tr}  \big(\widehat{f}(t)\widetilde{\bm{A}}(\bm{r}_1)\big)
\ast \langle \widetilde{\bm{A}}\overline{\bm{A}}\rangle_t^{-1}
(\bm{r}_1,\bm{r}_2) \nonumber \\
&&\qquad \ast \frac{1}{2} \Big( \overline{\bm{A}}(\bm{r}_2)\widehat{\rho}_L(t)
+ \hat{\rho}_L(t)\overline{\bm{A}}^\dagger(\bm{r}_2) \Big), \nonumber
\end{eqnarray}
and its complement $\widehat{\mathcal{Q}}^\dagger(t)=1-\widehat{\mathcal{P}}^\dagger(t)$. The symmetric form of the projector has been chosen to make its Hermitian structure explicit. This equation is expressed in notation that will be used in the remainder of this paper: $\langle \widetilde{\bm{A}}\overline{\bm{A}}\rangle_t^{-1}
(\bm{r}_1,\bm{r}_2)$ is the inverse of the overlap matrix $\langle \widetilde{\bm{A}}(\bm{r}_1) \overline{\bm{A}}(\bm{r}_2) \rangle_t$. An overline in $\overline{\bm{A}}(\bm{r})$ and its Hermitian conjugate $\overline{\bm{A}}^\dagger(\bm{r})$ is used to signify that they are nonequilibrium generalizations of Kubo-transformed variables.~\cite{kubo1957} More specifically, for any Hermitian operator $\widehat{O}(\bm{r})=\widehat{O}^\dagger(\bm{r})$, we define
\begin{equation}
\overline{O}(\bm{r}) = \int_0^1 dx \, e^{x\widehat{\bm{A}}\ast\bm{\phi}} \, \widetilde{O}(\bm{r})
e^{-x\widehat{\bm{A}}\ast\bm{\phi}},
\end{equation}
together with the corresponding expression for $\overline{O}^{\dagger}(\bm{r})$. Also, the presence of a tilde over an operator as in
\begin{equation}
\widetilde{O}(\bm{r}) = \widehat{O}(\bm{r})  - \langle \widehat{O}(\bm{r}) \rangle_t
\end{equation}
denotes the difference of the operator from its nonequilibrium average.

The projection operator formalism we use is based largely on that of Robertson~\cite{R66,R67}, which also been used to describe the dynamics of classical active particle systems~\cite{RSGK20,RSK24}. Related  time-dependent projectors are extensively discussed in Grabert~\cite{Grabert1977,grabert1982projection} . These projectors have been used in general formulations nonequilibrium statistical mechanics.~\cite{Ottinger2005}, and the work by Espa{\~n}ol~\cite{Espanol2023} is related to that presented here.

When the projector acts on the exact density $\widehat{\rho}(t)$, one finds
$\widehat{\mathcal{P}}^\dagger(t) \widehat{\rho}(t)=\widehat{\rho}_L(t)$, a consequence of the fact that
${\rm Tr}\big(\widehat{\rho}(t)\widetilde{\bm{A}}(\bm{r}_1)\big)=0$ holds by construction. Then, using the projector and its complement $\widehat{\mathcal{Q}}^\dagger(t)$ we can write $\widehat{\rho}(t) =\widehat{\rho}_L(t) +\widehat{\mathcal{Q}}^\dagger(t)\widehat{\rho}(t)$. Making use of the expression for $\widehat{\mathcal{Q}}^\dagger(t) \widehat{\rho}(t)$ derived in Appendix~\ref{app:projectors}, we obtain the following relation between the two density operators:
\begin{eqnarray}\label{eq:rho-rhoL}
\widehat{\rho}(t) &=& \widehat{\rho}_L(t)
    + \widehat{\mathcal{U}}^\dagger_Q(t,0)\widehat{\mathcal{Q}}^\dagger(0) \widehat{\rho}(0)  \\
    && -\int_0^t dt_1 \;
       \widehat{\mathcal{U}}^\dagger_Q(t,t_1) \widehat{\mathcal{Q}}^\dagger(t_1) i\widehat{\mathcal{L}} (\lambda_{t_1}) \hat{\rho}_L(t_1),\nonumber
\end{eqnarray}
where $\widehat{\mathcal{U}}^\dagger_Q(t,t_1)$ is the projected time-ordered exponential
\begin{equation}
\widehat{\mathcal{U}}^\dagger_Q(t,t_1)
    = \mathcal{T}_+ \exp\!\Big(-\int_{t_1}^t dt' \;
      \widehat{\mathcal{Q}}^\dagger(t')i\widehat{\mathcal{L}}(\lambda_{t'})\Big).
\end{equation}

Substitution of this expression for $\widehat{\rho}(t)$ into the last equality in Eq.~(\ref{eq:dadt}) results in an evolution equation where all averages are over the local density operator. After evaluating $i\widehat{\mathcal{L}}(\lambda_t) \widehat{\rho}_L(t)$ and moving the evolution operators from the density operator to the observables using ${\rm Tr} \big(\widehat{C}_1 \, \widehat{\mathcal{U}}^\dagger_Q(t,t_1) \widehat{C}_2(t_1) \big)={\rm Tr} \big( (\widehat{\mathcal{U}}_Q(t,t_1)\widehat{C}_1) \, \widehat{C}_2(t_1) \big)$, steps that are given in Appendix~\ref{app:av-value-eqs}, the resulting evolution equation is
\begin{eqnarray}\label{eq:Qa-phi}
\dot{\bm{a}}(\bm{r},t)&=& \langle  \widehat{\bm{\mathcal{J}}}_{A}(\bm{r}) \rangle_t + \bm{I}_A(\bm{r},t)  \\
&-& \int_0^t dt_1 \; \bm{\mathsf{L}}_{AA}(\bm{r},\bm{r}_1,t,t_1) \ast \bm{\phi}(\bm{r}_1,t_1) ,\nonumber
\end{eqnarray}
where the correlation function expression for the dissipative coefficient is written as
\begin{equation}\label{eq:L-gen}
\bm{\mathsf{L}}_{AA}(\bm{r},\bm{r}_1,t,t_1) = \big\langle  \widehat{\bm{{\mathcal J}}}_{A,t}(\bm{r},t_1,t)\overline{\bm{\mathcal{J}}}_{A,t_1}(\bm{r}_1)   \big\rangle^{\rm sym}_{t_1}.
\end{equation}
The time dependent flux is given by $\widehat{\bm{{\mathcal J}}}_{A,t}(\bm{r},t_1,t) =  \widehat{\mathcal{U}}_{Q}(t_1,t) \widehat{\bm{\mathcal{J}}}_{A,t}(\bm{r})$ with the projected flux operator defined by $\widehat{\bm{\mathcal{J}}}_{A,t}(\bm{r}) \equiv \widehat{\mathcal{Q}}(t)\widehat{\bm{\mathcal{J}}}_{A}(\bm{r})$ indicated by the subscript $t$. The adjoint of $\widehat{\mathcal{U}}_{Q}^\dagger(t_1,t)$ is $\widehat{\mathcal{U}}_{Q}(t_1,t)$.

The initial condition term $\bm{I}_A(t)={\rm Tr} \big( \widehat{\mathcal{U}}_Q(t,0)\widehat{\bm{\mathcal{J}}}_A)\widehat{\mathcal{Q}}^\dagger(0) \widehat{\rho}(0) \big)$ in the above equation will decay on a short time scale in view of the projected dynamics. Since this formulation is based on use of local density operators, at the initial time we take $\widehat{\rho}(0)$ to have a local form where $\widehat{\mathcal{Q}}^\dagger(0) \widehat{\rho}(0)=0$. For such an initial condition $\bm{I}_A$ is identically zero; henceforth, it will be omitted in the equations.

The equations of motion for the components of $\bm{a}(\bm{r},t)$ will be used extensively in what follows and we list them here: the subsystem average values satisfy
\begin{eqnarray}\label{eq:as}
&&\dot{\bm{a}}_s(t)=  \langle \widehat{\bm{\mathcal{J}}}_s \rangle_t  -\int_0^t dt_1 \; \Big[\bm{\mathsf{L}}_{ss}(t,t_1) \cdot  \bm{\phi}_s(t_1) \nonumber \\
&& \qquad \qquad \qquad +\bm{\mathsf{L}}_{s0}(\bm{r}_1,t,t_1)\ast  \phi_0(\bm{r}_1,t_1)\Big]
\end{eqnarray}
with the following explicit expressions for the dissipative coefficients: $\bm{\mathsf{L}}_{ss}(t,t_1) =\langle  \widehat{\bm{\mathcal{J}}}_{s,t}(t_1,t))) \overline{\bm{\mathcal{J}}}_{s,t_1} \rangle^{\rm sym}_{t_1}$ and $\bm{\mathsf{L}}_{s0}(\bm{r}_1,t,t_1)=\langle  \widehat{\bm{\mathcal{J}}}_{s,t}(t_1,t))) \overline{{\mathcal J}}_{0,t_1}(\bm{r}_1) \rangle_{t_1}$. The reservoir average energy density obeys
\begin{eqnarray}\label{eq:e0}
&&\dot{e}_0(\bm{r},t)=  \langle \widehat{{\mathcal J}}_0(\bm{r}) \rangle_t  -\int_0^t dt_1  \Big[  \bm{\mathsf{L}}_{0s}(\bm{r},t,t_1)\cdot  \bm{\phi}_s(t_1) \nonumber \\
&& \qquad \qquad \qquad  \qquad +  {\mathsf{L}}_{00}(\bm{r},\bm{r}_1,t,t_1)\ast  \phi_0(\bm{r}_1,t_1)\Big]
\end{eqnarray}
with coefficients $ \bm{\mathsf{L}}_{0s}(\bm{r},t,t_1)=\langle  \widehat{{\mathcal J}}_{0,t}(\bm{r},t_1,t))) \overline{\bm{\mathcal{J}}}_{s,t_1} \rangle^{\rm sym}_{t_1}$ and ${\mathsf{L}}_{00}(\bm{r},\bm{r}_1,t,t_1)=\langle  \widehat{{\mathcal J}}_{0,t}(\bm{r},t_1,t))) \overline{{\mathcal J}}_{0,t_1}(\bm{r}_1) \rangle^{\rm sym}_{t_1}$.

The fluxes in these equations are given by $\widehat{\bm{\mathcal{J}}}_s =\frac{i}{\hbar} [\widehat{\mathsf{H}}(\lambda_t), \widehat{\bm{A}}_s]$ and $\widehat{\mathcal{J}}_0(\bm{r}) =\frac{i}{\hbar} [\widehat{\mathsf{H}}(\lambda_t), \widehat{H}_0(\bm{r})]$. The expressions for the components of $\widehat{\bm{\mathcal{J}}}_s=(\widehat{\mathcal{J}}_B,\widehat{\mathcal{J}}_S)$ are
\begin{eqnarray}\label{eq:fluxBandS}
\widehat{{\mathcal{J}}}_{B} &=&\frac{i}{\hbar} [\widehat{H}_S,\widehat{B}]  \\
\widehat{{\mathcal{J}}}_{S} &=& \frac{i}{\hbar} [\widehat{V}_c, \widehat{H}_S]- \frac{i}{\hbar} [\widehat{H}_S, \widehat{B}]\lambda_t \equiv -\widehat{J}_{c} - \widehat{{\mathcal{J}}}_{B} \lambda_t \nonumber
\end{eqnarray}
while the flux of $\widehat{H}_0(\bm{r})$ is
\begin{eqnarray}\label{eq:flux0}
\widehat{{\mathcal{J}}}_{0}(\bm{r}) &=&\frac{i}{\hbar} [\widehat{H}_S, \widehat{V}_c(\bm{r})]+\frac{i}{\hbar} [\widehat{H}_0, \widehat{H}_0(\bm{r})]  \\
&\equiv& \widehat{J}_c(\bm{r}) -\partial_{\bm{r}} \cdot \widehat{\bm{j}}_0(\bm{r}). \nonumber
\end{eqnarray}
Since $\int d\bm{r} \; [\widehat{H}_0, \widehat{H}_0(\bm{r})]=0$, we have defined the flux operator $\widehat{\bm{j}}_0(\bm{r})$ by the relation $\frac{i}{\hbar} [\widehat{H}_0, \widehat{H}_0(\bm{r})]= -\partial_{\bm{r}} \cdot \widehat{\bm{j}}_0(\bm{r})$ given in the last equality in the equation.~\cite{local-operators}

\subsection{Constraint functions}\label{subsec:constraint}
The evolution equations for the average fields $\bm{a}(\bm{r},t)$ are not closed since they are expressed in terms of the constraint fields $\bm{\phi}(\bm{r},t)$ that remain to be determined. Another relation between these two fields can be found by taking the time derivative of the local density to obtain,
\begin{eqnarray}\label{eq:drhoL-dt}
\dot{\widehat{\rho}}_L(t) = \big( \overline{\bm{A}}(\bm{r})  \ast \dot{\bm{\phi}}(\bm{r},t) \big) \widehat{\rho}_L(t).
\end{eqnarray}
Then, using this expression to evaluate the time derivative of the average field $\bm{a}(\bm{r},t) ={\rm Tr} \big(\widehat{\bm{A}}(\bm{r}) \hat{\rho}_L(t) \big)$, we get
\begin{equation}\label{eq:adot-phidot}
\dot{\bm{a}}(\bm{r},t)=\langle \widetilde{\bm{A}}(\bm{r}) \overline{\bm{A}}(\bm{r}_1)\rangle_t \ast \dot{\bm{\phi}}(\bm{r}_1,t).
\end{equation}

More specifically for our operators, $\{\widehat{\bm{A}}(\bm{r}) \} = \{ \widehat{\bm{A}}_s, \widehat{A}_0 (\bm{r}) \}$, we can write the coupled equations,
\begin{eqnarray}
\dot{\bm{a}}_s(t)&= &  \langle \widetilde{\bm{A}}_s \overline{\bm{A}}_s)\rangle_t \cdot  \dot{\bm{\phi}}_s(t)  \label{eq:dtas} \\
&& \qquad  \quad
+\langle \widetilde{\bm{A}}_s \overline{A}_0(\bm{r}_1)\rangle_t \ast   \dot{\phi}_0(\bm{r}_1,t), \nonumber \\
\dot{e}_0(\bm{r},t)&=&  \langle \widetilde{A}_0(\bm{r}) \overline{\bm{A}}_s\rangle_t \cdot  \dot{\bm{\phi}}_s(t) \label{eq:dte0} \\
&& \qquad  \quad +\langle \widetilde{A}_0(\bm{r}) \overline{A}_0(\bm{r}_1)\rangle_t \ast  \dot{\phi}_0(\bm{r}_1,t). \nonumber
\end{eqnarray}
These equations can be used in Eqs.~(\ref{eq:as}) and (\ref{eq:e0}) to construct a set of coupled closed equations for the $\bm{\phi}(t)$ and $\phi_0(\bm{r},t)$ constraints.

Letting $\bm{K}_t(\bm{r},\bm{r}_1)=\langle \widetilde{\bm{A}}(\bm{r}) \overline{\bm{A}}(\bm{r}_1)\rangle_t$, we can also write the general equation for the $\bm{\phi}$ constraints as,
\begin{eqnarray}\label{eq:Qphi}
&&\bm{K}_t(\bm{r},\bm{r}_1) \ast \dot{\bm{\phi}}(\bm{r}_1,t) =  \langle \hat{\bm{\mathcal{J}}}_A(\bm{r}) \rangle_t \\
&&\qquad \quad -\int_0^t dt_1 \; \bm{\mathsf{L}}_{AA}(\bm{r},\bm{r}_1,t,t_1) \ast \bm{\phi}(\bm{r}_1,t_1). \nonumber
\end{eqnarray}
The equation written in this form will be useful in some of the calculations presented below.

\subsection{Entropy production}\label{subsec:entropy-prod}
The time rate of change of the entropy of the entire system can be obtained by differentiating the entropy definition in Eq.~(\ref{eq:entropy}):
\begin{eqnarray} \label{eq:dSdt-gen}
&&\dot{S}(t)=-k_B {\rm Tr} \big(\dot{\widehat{\rho}}_L(t)\big)\big(\widehat{\bm{A}}(\bm{r}) \ast \bm{\phi}(\bm{r},t)\big)  =\\
&&-k_B \bm{\phi}(\bm{r},t) \ast \langle \widetilde{\bm{A}}(\bm{r}_1) \overline{\bm{A}}(\bm{r})\rangle_t \ast \dot{\bm{\phi}}(\bm{r},t)
= -k_B \bm{\phi} \ast \dot{\bm{a}} \nonumber
\end{eqnarray}
where Eq.~(\ref{eq:drhoL-dt}) was used to obtain the second equality, followed by use of Eq.~(\ref{eq:adot-phidot}) to get the last equality. The equation of motion~(\ref{eq:Qa-phi}) for $\bm{a}(\bm{r},t)$ can then be used to write the following expression for the entropy production:
\begin{equation}\label{eq:S-prod}
 \sigma(t)\equiv k_B \bm{\phi}(\bm{r},t) \ast
\int_0^t dt_1 \; \bm{\mathsf{L}}_{AA}(\bm{r},\bm{r}_1,t,t_1) \ast \bm{\phi}(\bm{r}_1,t_1).
\end{equation}

The nonequilibrium average fluxes, $\langle \bm{\mathcal{J}}_A \rangle_t$, do not appear in the entropy production since the sum of the these contributions vanishes. This follows from the computation,
\begin{eqnarray}\label{eq:avJaphi=0}
&&\langle \bm{\mathcal{J}}_A \rangle_t \ast \bm{\phi}= {\rm Tr}\Big( \big(i\widehat{\mathcal{L}}(\lambda_t) \widehat{\bm{A}}\ast \bm{\phi}\big) \widehat{\rho}_L(t)\Big)\\
&& \qquad \quad =-{\rm Tr}\Big( \big(\widehat{\bm{A}} \ast \bm{\phi} \big) \big(i\widehat{\mathcal{L}}(\lambda_t)\widehat{\rho}_L(t)\big)\Big)\nonumber \\
&&\qquad \quad = -\frac{i}{\hbar}{\rm Tr}\Big( \big(\widehat{\bm{A}} \ast \bm{\phi} \big) [\mathsf{H}(\lambda_t),e^{\widehat{\bm{A}} \ast \bm{\phi}}]\Big)/\mathcal{Z}(t)=0.\nonumber
\end{eqnarray}
After moving the evolution operator onto the density in the second line and inserting its definition in terms of the commutator in the third line, its value of zero in the last line follows after using cyclic permutations under the trace. The entropy production will be discussed in more detail in the next section.

\section{Internal energy and entropy}\label{sec:U}

A main thermodynamic quantity is the internal energy $U$ of the quantum subsystem, and this section deals with its microscopic dynamics and the thermodynamic relations it satisfies. The microscopic definition of the average internal energy at time $t$ is given by the expectation value of the subsystem Hamiltonian $\widehat{H}_S(\lambda_t)$ in the presence of external driving that may do work on the subsystem:
$U(t)={\rm Tr} \big( \widehat{H}_S(\lambda_t) \widehat{\rho}(t) \big)$. Since the Hamiltonian depends explicitly on the time through the autonomous variable $\lambda_t$, the total time derivative of the internal energy is
\begin{eqnarray}\label{eq: dUdt}
\dot{U}(t)&=& {\rm Tr} \big(\partial_t \widehat{H}_S(\lambda_t) \big) \widehat{\rho}(t) +{\rm Tr}  \widehat{H}_S(\lambda_t) \partial_t \widehat{\rho}(t) \nonumber \\
&\equiv& {\rm Tr} \widehat{w}_t \widehat{\rho}(t) +{\rm Tr}  \widehat{\mathcal{J}}_U \widehat{\rho}(t).
\end{eqnarray}
The first term on the right accounts for the temporal variation of $U(t)$ due to the explicit time dependence of the Hamiltonian, while the second term arises from the evolution of the full system density matrix. In the second line we introduced the shorthand notation, $\widehat{w}_t= \partial_t \widehat{H}_S(\lambda_t)$ and for our choice of subsystem Hamiltonian this operator can also be written as $\widehat{w}_t=\dot{\lambda}_t \widehat{B}$. Also in the second line, the term arising from the time derivative of the density operator is rewritten in terms of the internal energy flux,
\begin{equation}\label{eq:JU-Jc}
\hat{\mathcal{J}}_U= \frac{i}{\hbar}[\widehat{\mathsf{H}}(\lambda_t),\widehat{H}_S(\lambda_t)] =-\frac{i}{\hbar}[\widehat{H}_S,\hat{V}_c] =-\widehat{J}_c,
\end{equation}
after using the quantum Liouville equation for the time derivative of the density operator, and moving the Liouville operator onto the subsystem Hamiltonian.

Since $\widehat{w}_t=\dot{\lambda}_t \widehat{B}$, in view of the constraint conditions we can write its expectation value over the full density operator equivalently as an average over the local density operator:   ${\rm Tr} \widehat{w}_t \widehat{\rho}(t)= {\rm Tr} \widehat{w}_t \widehat{\rho}_L(t) = \langle \widehat{w}_t \rangle_t$. The same replacement cannot be made in ${\rm Tr}  \widehat{\mathcal{J}}_U \widehat{\rho}(t)$ because the flux $\widehat{\mathcal{J}}_U$ is not in the list of constrained operators. Instead, one needs to use the relation between the full and local density operators in Eq.~(\ref{eq:rho-rhoL}) to write this average value in terms of the local density operator.

An alternative way to compute $\dot{U}(t)$ is to differentiate its definition, $U(t)= e_S(t) +\lambda_t b(t)$, to obtain
\begin{eqnarray}
\dot{U}(t)- \langle w_t \rangle_t &=&\lambda_t \dot{b}(t) +\dot{e}_S(t)\\
&=&(\lambda_t, 1)\cdot \dot{\bm{a}}_s(t),\nonumber
\end{eqnarray}
and use Eq.~(\ref{eq:as}) to compute the time derivatives of the average $\bm{a}_s(t)$ fields. Carrying out these steps we find
\begin{eqnarray}\label{eq:U2}
&&\dot{U}(t)- \langle \hat{w}_t \rangle_t =  \langle \widehat{{\mathcal J}}_U \rangle_t  \\
&& +\int_0^t dt_1 \; \Big[ {\mathsf{L}}_{UB}(t,t_1) \gamma(t_1)+ {\mathsf{L}}_{UU}(t,t_1)\beta_S(t_1) \nonumber \\
&& \qquad  \qquad  \quad +{\mathsf{L}}_{U0}(\bm{r}_1,t,t_1)\ast  \beta_0(\bm{r}_1,t_1)\Big], \nonumber
\end{eqnarray}
with definitions for the dissipative coefficients similar  to those defined earlier: ${\mathsf{L}}_{UB}(t,t_1)=\langle  \widehat{{\mathcal J}}_{U,t}(t_1,t)) \overline{{\mathcal J}}_{B,t_1} \rangle^{\rm sym}_{t_1}$, ${\mathsf{L}}_{UU}(t,t_1)=\langle  \widehat{{\mathcal J}}_{U,t}(t_1,t)) \overline{{\mathcal J}}_{U,t_1} \rangle^{\rm sym}_{t_1}$ and ${\mathsf{L}}_{U0}(\bm{r}_1,t,t_1)=\langle  \widehat{{\mathcal J}}_{U,t}(t_1,t))) \overline{{\mathcal J}}_{0,t_1}(\bm{r}_1) \rangle^{\rm sym}_{t_1}$. The combination of constraint variables,
\begin{equation} \label{eq:gamma-def}
\gamma(t)=\alpha(t) -\lambda_{t} \beta_S(t)
\end{equation}
in this equation arises from the relations among the fluxes given in Eq.~(\ref{eq:fluxBandS}). In Eq.~(\ref{eq:U2}) the contribution $\dot{U}(t)- \langle \hat{w}_t \rangle_t$ is the sum of the nonequilibrium average flux, $\langle \widehat{{\mathcal J}}_U \rangle_t$, and dissipative terms that depend on nonequilibrium flux-flux correlation functions.

It is useful to note several additional features of the components of these equations. From Eq.~(\ref{eq:JU-Jc}), $\hat{\mathcal{J}}_U=-(i/\hbar)[\widehat{H}_S,\widehat{V}_c]=-\widehat{J}_c$, one can see that the internal energy flux is confined to the coupling region where the quantum subsystem interacts with the reservoir. Under equilibrium conditions the average flux $\langle \widehat{{\mathcal J}}_U \rangle_t$ vanishes but under nonequilibrium conditions this term gives rise to a non-dissipative mean flux that can change the internal energy of the subsystem. Also, the flux $\widehat{{\mathcal{J}}}_{0}(\bm{r})$ in Eq.~(\ref{eq:flux0}) has two components: $\widehat{J}_c(\bm{r})$ that is confined to the coupling region, and $\partial_{\bm{r}} \cdot \widehat{\bm{j}}_0(\bm{r})$ that is dominated by reservoir contributions. The flux $\mathcal{J}_B=(i/\hbar)[\widehat{H}_S,\widehat{B}]$ in Eq.~(\ref{eq:fluxBandS}) is confined to the quantum subsystem. Depending on the physical context, the external driving operator $\widehat{B}$ may couple to either the entire subsystem or only the portion that does not couple to the reservoirs. We assume the latter case for simplicity so that the ${\mathsf{L}}_{UB}$ (and ${\mathsf{L}}_{BU}$ )dissipative coefficients can be dropped. In this case $b(t)$ satisfies,
\begin{equation}\label{eq:b}
\dot{b}(t)= \langle \widehat{\mathcal{J}}_B\rangle_t +\int_0^t dt_1 \; {\mathsf{L}}_{BB}(t,t_1)\gamma(t_1).
\end{equation}

On the basis of these considerations, we can write evolution equation for the internal energy~(\ref{eq:U2}) in a form that better displays the nature of the coupling terms:
\begin{eqnarray}\label{eq:U2-trans}
&&\dot{U}(t)- \langle \hat{w}_t \rangle_t =  -\langle \widehat{J}_ c\rangle_t +\int_0^t dt_1 \; \Big[{\mathsf{L}}_{cc}(t,t_1)\beta_S(t_1) \nonumber \\
&&\qquad \qquad \quad -{\mathsf{L}}_{cc}(\bm{r}_1,t,t_1)\ast  \beta_0(\bm{r}_1,t_1) \nonumber \\
&&\qquad \qquad \quad - \bm{\mathsf{L}}_{c0}(\bm{r}_1, t,t_1)\ast \partial_{\bm{r}_1} \beta_0(\bm{r}_1,t_1)\Big].
\end{eqnarray}
All of the terms on the right side of this equation, which can be identified as ${\rm Tr}  \widehat{\mathcal{J}}_U \widehat{\rho}(t)$, are either confined to the coupling region or characterize the coupling to the reservoirs through the interaction region.

Integrating the internal energy evolution equation, $\dot{U}(t)= \langle \widehat{w}_t \rangle_t + {\rm Tr}  \widehat{\mathcal{J}}_U \widehat{\rho}(t)$, we find the first law of thermodynamics,
\begin{equation}
\Delta U(t) =U(t)-U(0)=W(t) + Q(t),
\end{equation}
with work $W(t)$ and heat $Q(t)$, respectively,  given by
\begin{equation}\label{eq:work-heat}
W(t)=\int_0^t dt_1 \;\langle \widehat{w}_{t_1} \rangle_{t_1},  \; Q(t)=\int_0^t dt_1 \; {\rm Tr} \widehat{\mathcal{J}}_U \widehat{\rho}(t_1).
\end{equation}
It is also instructive to write Eq.~(\ref{eq:U2}) in an equivalent form by noting that if one integrates Eq.~(\ref{eq:e0}) over all space, the resulting equation for $\dot{e}_0(t)$ is just given by the negative of the right side of Eq.~(\ref{eq:U2}), so that
\begin{equation}
\dot{U}(t)- \langle \hat{w}_t \rangle_t = -\dot{e}_0(t),
\end{equation}
and the internal energy change can now be written as
\begin{equation}
\Delta U(t) =U(t)-U(0)=W(t) -\Delta e_0(t),
\end{equation}
so that $Q(t)=-\Delta e_0(t)=-(e_0(t)-e_0(0))$ is just equal to negative of the energy change in the reservoir. As expected, it is clear that the work and heat in these expressions are not state functions since the time integrals depend on the protocol used for $\lambda_t$.

The development presented thus far has provided formally exact, nonlinear, general evolution equations for the $b$, $U$ and $e_0$ variables. The equations describe how the internal energy of a quantum system with unitary evolution changes when it is subject to external driving and coupled to nonequilibrium reservoirs. However, the equations are difficult to analyze and implement due to the complexity of the kernels that involve both memory and spatial non-locality. The equations of motion adopt a more transparent and computationally tractable form if physically-motivated approximations to the dissipative kernels are made. In particular, we shall make use of Markovian and localization approximations to these kernels.

For conserved (or nearly conserved) variables such as those considered here, the correlation functions of their fluxes will typically decay on a faster time scale than that for the average fields and constraint functions. In this case the memory kernels can be approximated by their Markovian form, $\bm{\mathsf{L}}_{AA}(\bm{r},\bm{r}_1,t,t_1) \approx \bm{\mathsf{L}}_{AA}(\bm{r},\bm{r}_1,t)$, where the dissipative coefficient is defined by
\begin{equation}\label{eq:Mark-gen}
\bm{\mathsf{L}}_{AA}(\bm{r},\bm{r}_1,t) = \int_0^\infty dt_1 \; \big\langle  \widehat{\bm{{\mathcal J}}}_{A,t}(\bm{r},t_1)\overline{\bm{\mathcal{J}}}_{A,t}(\bm{r}_1)   \big\rangle^{\rm sym}_{t}.
\end{equation}
Depending on the physical context, the Markovian approximation to the dissipative coefficients may be appropriate for the entire set of variables or only a portion of the set; for instance, it may only be appropriate for the dissipative processes in the reservoir.

Furthermore, the flux-flux correlation function $\bm{\mathsf{L}}_{AA}(\bm{r},\bm{r}_1,t)$ will be vanishingly small for distances $|\bm{r}-\bm{r}_1|$ greater than a small microscopic length, so we can make use of a localization approximation and write

\begin{equation}
\bm{\mathsf{L}}_{AA}(\bm{r},\bm{r}_1,t) \ast \bm{\phi}(\bm{r}_1,t) \approx \bm{\mathsf{L}}_{AA}(\bm{r},t) \bm{\phi}(\bm{r},t),
\end{equation}
where $\bm{\mathsf{L}}_{AA}(\bm{r},t)=\langle  \widehat{\bm{{\mathcal J}}}_{A,t}(\bm{r},t)\overline{\bm{\mathcal{J}}}_{A,t}   \rangle^{\rm sym}_{t}$.

As an example, if we make a Markovian approximation to Eq.~(\ref{eq:U2-trans}), the dissipative contributions can be written as:
\begin{eqnarray}
&&\int_0^t dt_1  {\mathsf{L}}_{cc}(t,t_1)\beta_S(t_1) \approx {L}_{cc}(t)\beta_S(t) \\
&&\int _0^t dt_1 {\mathsf{L}}_{cc}(\bm{r}_1,t,t_1)\ast  \beta_0(\bm{r}_1,t_1)\approx L_{cc}(\bm{r}_1,t)\ast  \beta_0(\bm{r}_1,t),\nonumber \\
\end{eqnarray}
where
\begin{eqnarray}
L_{cc}(t)&=&\int_0^\infty  dt_1 \;  \langle  \widehat{J}_{c,t}(t_1,t)) \overline{J}_{c,t} \rangle^{\rm sym}_{t}, \label{eq:Lcc-coef} \\
L_{cc}(\bm{r}_1,t)&=&\int_0^\infty  dt_1 \;  \langle  \widehat{J}_{c,t}(t_1,t)) \overline{J}_{c,t}(\bm{r}_1) \rangle^{\rm sym}_{t}.\label{eq:Lcc-coef-r}
\end{eqnarray}
We shall not consider the contribution $\bm{\mathsf{L}}_{c0}$ since this dissipative coefficient involves a correlation function with fluxes of different tensorial rank. The expression in Eq.~(\ref{eq:Lcc-coef-r}) is nonlocal in space but from the structure the kernel in this equation one can see that only $\bm{r}_1$ values in the vicinity of the coupling zone will contribute to the correlation function since the $\widehat{J}_c$ fluxes are localized to this region. In this circumstance, we can replace $\beta_0(\bm{r},t)$ by its typical value in the coupling region, $\beta_0(\bm{r},t) \approx \beta_c(t)$, and use of this condition gives
\begin{equation}
L_{cc}(\bm{r}_1,t)\ast  \beta_0(\bm{r}_1,t) \approx L_{cc}(t)  \beta_c(t).
\end{equation}
After these approximations are applied, the internal energy equation takes the simpler form,
\begin{eqnarray}\label{eq:U2-approx}
\dot{U}(t)- \langle \hat{w}_t \rangle_t = - \langle \widehat{J}_c \rangle_t +{L}_{cc}(t)\big(\beta_S(t) - \beta_c(t)\big),
\end{eqnarray}
and the Markovian approximation for the $b(t)$ evolution equation is
\begin{equation}\label{eq:b2}
\dot{b}(t)= \langle \widehat{\mathcal{J}}_B\rangle_t + {\mathsf{L}}_{BB}(t)\gamma(t).
\end{equation}

\subsection{Entropy production}
Writing Eq.~(\ref{eq:S-prod}) for the entropy production in the system for the variables $b$, $U$ and $e_0$ we have
\begin{eqnarray}\label{eq:S-change-2}
\sigma(t)&=& k_B \int_0^t  dt_1 \Big[ \gamma(t) \big(\mathsf{L}_{BB}(t,t_1)\gamma(t_1) \big)  \\
&+& \beta_S(t) \big(\mathsf{L}_{UU}(t,t_1)  \beta_S(t_1) \nonumber \\
&& \qquad \qquad  +\mathsf{L}_{U0}(\bm{r}_1,t,t_1)\ast  \beta_0(\bm{r}_1,t_1) \big) \nonumber \\
&+&\beta_{0}(\bm{r},t) \ast \big(\mathsf{L}_{0U}(\bm{r},t,t_1)  \beta_S(t_1) \nonumber\\
&& \qquad  \qquad  +\mathsf{L}_{00}(\bm{r},\bm{r}_1,t,t_1)\ast  \beta_0(\bm{r}_1,t_1)  \big)\Big].\nonumber
\end{eqnarray}
Following the discussion in Sec.~\ref{subsec:entropy-prod}, the non-dissipative average fluxes do not contribute to the entropy production since $\langle \widehat{{\mathcal J}}_B \rangle_t\gamma(t_1)+\langle \widehat{{\mathcal J}}_U \rangle_t \beta_S(t)+ \langle \widehat{{\mathcal J}}_0(\bm{r}) \rangle_t \ast \beta_0(\bm{r})=0$.

It is difficult to draw conclusions about the sign of $\sigma$ due to the complexity of the dissipative coefficients and memory as discussed above. If we implement the same Markovian and localization approximations as in the internal energy equation, we obtain,
\begin{eqnarray}\label{eq:sigma-mark}
\sigma(t)&=& k_B \Big[ \gamma(t) {L}_{BB}(t)\gamma(t)   \\
&+& \big(\beta_S(t) -\beta_c(t) \big) {L}_{cc}(t) \big(  \beta_S(t) - \beta_c(t) \big) \nonumber\\
&+&(\partial_{\bm{r}}\beta_{0}(\bm{r},t))\cdot \bm{\mathsf{L}}_{00}(\bm{r},t)\cdot  (\partial_{\bm{r}}\beta_0(\bm{r},t))\Big] \ge 0,\nonumber
\end{eqnarray}
with
\begin{equation}\label{eq:L00}
\bm{\mathsf{L}}_{00}(\bm{r}, t)=\int_0^\infty dt_1 \; \big\langle  \widehat{\mathbf{j}}_{0,t}(\bm{r},t_1)\overline{\mathbf{j}}_{0,t}  \big\rangle^{\rm sym}_{t}.
\end{equation}
Equation~(\ref{eq:sigma-mark}) shows that the entropy production is the sum of three terms corresponding to contributions from the quantum subsystem, the interfacial region, and the reservoirs, respectively, each of which is separately positive semi-definite. We can identify the following thermodynamic forces $X$ and fluxes $\mathsf{J}$:  $X_\gamma=k_B\gamma(t)$ and $\mathsf{J}_\gamma(t)={L}_{BB}(t)\gamma(t)$ for the quantum subsystem, $X_c=k_B\big(\beta_S(t) -\beta_c(t)\big)$,  $\mathsf{J}_c={L}_{cc}(t) \big(  \beta_S(t) - \beta_c(t) \big)$ in the coupling region and $X_0=k_B\partial_{\bm{r}}\beta_{0}(\bm{r},t)$ and $\mathsf{J}_0=\bm{\mathsf{L}}_{00}(\bm{r},t)\cdot  (\partial_{\bm{r}}\beta_0(\bm{r},t))$ in the reservoir. The expression for the entropy production can then be written in standard form, $\sigma=\sum_i \mathsf{J}_i X_i$.

While this expression for $\sigma$ has an apparently bilinear form, since the average values in the dissipative coefficients are taken over the local densities, the entropy production is a highly nonlinear function of the thermodynamic forces. Only in the limit where the dissipative coefficients are approximated by replacing the nonequilibrium local density by its equilibrium form do they satisfy Onsager reciprocal relations and is $\sigma$ strictly bilinear If the nonequilibrium local density is expanded in a power series of the thermodynamic forces $\bm{\phi}$, higher order Onsager-like contributions can be identified.

\section{Subsystem with driving in the absence of coupling to the reservoirs}
We now suppose that the quantum subsystem is isolated so that there is no coupling to the reservoirs
and thus no heat transfer. We then have $\widehat{{\mathcal{J}}}_U =0$, so that Eq.~(\ref{eq:U2}) reduces
to the expected form,
\begin{equation}
\dot{U}(t)= \langle \widehat{{w}}_{t} \rangle_t =\langle \widehat{{w}}_t \rangle(t).
\end{equation}
Because the average value of $\widehat{{w}}_{t}=\dot{\lambda}_t \widehat{B}$ is given exactly by either the local density average or the average over the full density operator, ${\rm Tr} \widehat{w}_t \widehat{\rho}(t) \equiv \langle \widehat{w}_t \rangle(t)$, we can compute its value using either expression. For external driving that takes the system from $\lambda_0$ to $\lambda_t$ we have
\begin{equation}
\Delta U(t)=\int_0^t dt_1 \langle \widehat{{w}}_{t_1} \rangle(t_1)
\equiv  W(t),
\end{equation}
which is the usual relation expressing the change in the average internal energy as the work done on the system in the process. If $\lambda_t$ does not change but remains at $\lambda_0$, no work is done, and we have the expected result $\Delta U(t)=0$ since energy is conserved.

The entropy production for the isolated system can be obtained from $\sigma$ in Eq.~(\ref{eq:sigma-mark}) by setting the coupling flux $J_c=0$ so that $\sigma$ reduces to the sum of two independent contributions for the isolated subsystem and reservoir. The isolated subsystem entropy production is
\begin{equation}\label{eq:sigma-adiab-b}
\sigma_\gamma(t)= k_B \gamma(t) \mathsf{L}_{BB}(t)\gamma(t) \ge 0,
\end{equation}
while the isolated reservoir contribution will be discussed in the next section. From this expression we see that the entropy production $\sigma_\gamma$ arises from dissipative processes related to external driving operator $\widehat{B}$ variable that is included in the set of subsystem variables. This is a consequence of our choice of operators in the set $\bm{A}$ that was motivated by the properties of the quantum subsystem coupled to reservoirs we wished to investigate. In particular, by choosing to describe the isolated quantum subsystem without external driving by the single operator $\widehat{H}_S$ in the set, its evolution is unitary. To describe internal dissipation processes requires the addition of other subsystem operators, e.g. populations, corresponding to a finer coarse graining, which we have chosen not to do in this work.

\subsection{A specific work protocol}
 Suppose that the autonomous parameter $\lambda_t$ in the Hamiltonian consists of a sequence of $m$ segments of length $\tau$ where $\lambda_t$ takes the values $\{\lambda_0, \lambda_0, \dots, \lambda_m=\lambda_f \}$. The initial and final values of $\lambda$ that characterize the initial and final system states are fixed at $\lambda_0$ and $\lambda_f$, respectively, while the intermediate values taken depend on $m$ and $\tau$. (A similar protocol was used by Crooks~\cite{Crooks1998}.) For this protocol we have,
\begin{equation}
\lambda_t= \sum_{n=0}^m \lambda_n H(t-n\tau_+)H((n+1)\tau_- -t),
\end{equation}
where the Heaviside function $H(x)= 1$ for $x\ge 0$, and zero otherwise, and $n \tau_{\pm}= n \tau \pm \epsilon$ where $\epsilon$ is an infinitesimal quantity. The time derivative of $\lambda_t$ is
\begin{equation}
\dot{\lambda}_t= \sum_{n=1}^m (\lambda_n-\lambda_{n-1}) \delta(t-n\tau).
\end{equation}
Substituting these expressions into the formula for the work,
\begin{equation}
W(t_f)=\int_0^{t_f} dt \;{\rm Tr} \big(\partial_t \widehat{H}_S(\lambda_t)\big) \widehat{\rho}(t),
\end{equation}
where $t_f$ is the time required to reach the final state for a given protocol, we obtain,
\begin{eqnarray}
W(t_f)&=& \sum_{n=1}^m {\rm Tr}\big( \widehat{H}_S(\lambda_n)- \widehat{H}_S(\lambda_{n-1}) \big) \widehat{\rho}(n\tau) \\
&=&\sum_{n=1}^m {\rm Tr}\big( \widehat{H}_S(\lambda_n)- \widehat{H}_S(\lambda_{n-1}) \big)\widehat{\mathcal{U}}^\dagger(n\tau,0)\widehat{\rho}(0). \nonumber
\end{eqnarray}
The evolution operator is a concatenation of segments,
\begin{equation}
\widehat{\mathcal{U}}^\dagger(n\tau,0)=\widehat{\mathcal{U}}^\dagger(n\tau,(n-1)\tau)  \dots \widehat{\mathcal{U}}^\dagger(2\tau,\tau) \widehat{\mathcal{U}}^\dagger(\tau,0),
\end{equation}
where the evolution operators in each segment specify evolution governed by a Hamiltonian with a specific $\lambda$ value: $\widehat{\mathcal{U}}^\dagger(n\tau,(n-1)\tau)= \exp{(-\widehat{\mathcal{L}}(\lambda_{n-1})\tau})$. In this time interval the Hamiltonian is $\widehat{H}_S(\lambda_{n-1})$. We can then write
\begin{eqnarray}
&&{\rm Tr}\widehat{H}_S(\lambda_{n-1}) \big)\widehat{\mathcal{U}}^\dagger(n\tau,0)\widehat{\rho}(0)\\
&& \quad ={\rm Tr}\big(\widehat{\mathcal{U}}(n\tau,(n-1)\tau)\widehat{H}_S(\lambda_{n-1})\big)
\widehat{\mathcal{U}}^\dagger((n-1)\tau,0)\widehat{\rho}(0)\nonumber \\
&& \quad = {\rm Tr}\widehat{H}_S(\lambda_{n-1})\widehat{\mathcal{U}}^\dagger((n-1)\tau,0)\widehat{\rho}(0),\nonumber
\end{eqnarray}
since $\widehat{H}_S(\lambda_{n-1})$ is conserved in this time interval.

Substituting this result in the work expression, we have
\begin{eqnarray}
W(t_f)&=& \sum_{n=1}^m {\rm Tr}\big(\widehat{\mathcal{U}}(n\tau,0) \widehat{H}_S(\lambda_n)\\
&& \qquad - \widehat{\mathcal{U}}((n-1)\tau,0)\widehat{H}_S(\lambda_{n-1}) \big)\widehat{\rho}(0), \nonumber
\end{eqnarray}
and the sum yields the expected result
\begin{eqnarray}
W(t_f)&=& {\rm Tr}\big( \widehat{\mathcal{U}}(m\tau,0)\widehat{H}_s(\lambda_m)\big)\widehat{\rho}(0)- {\rm Tr}\widehat{H}_s(\lambda_0) \widehat{\rho}(0)\nonumber \\
&=& U(t_m)-U(0).
\end{eqnarray}

A calculation that gives the same result can be carried out starting with the local density average, and this calculation is given in Appendix~\ref{app:adiab-local}.

\subsection{Reversible dynamics}
From the work expressions given above it follows that the magnitude of $W(t_f)$ depends on the protocol used to move from $\lambda_0$ to $\lambda_f$; i.e., it depends on the parameters, $\Delta \lambda_n=\lambda_n-\lambda_{n-1}$, the number $m$ of steps to reach $\lambda_f$ and the time interval $\tau$ between steps. As in standard discussions of work in thermodynamics, reversible dynamics may be obtained from such a protocol by taking the increments $\Delta \lambda_n \to 0$ while the number of steps $m \to \infty$ with $\lambda_m=\lambda_f$. The time intervals $\tau$ should be long enough for the system to relax to a (quasi-)equilibrium state at each interval.

We now discuss the conditions under which the local density,
\begin{equation}\label{eq:rhoL-alpha-beta}
\widehat{\rho}_L (t) = e^{-\widehat{H}_S \beta_S(t) -\widehat{B} \alpha(t)}/\mathcal{Z}(t),
\end{equation}
will evolve to the equilibrium density,
\begin{equation}\label{eq:rho-eq-n}
\widehat{\rho}_{\rm eq} = e^{-\widehat{H}(\lambda_n) \beta}_S/\mathcal{Z}_{\rm eq},
\end{equation}
in the $n^{th}$ time interval $\tau$ where the automatous driving parameter has value $\lambda_n$. The local density operator in Eq.~(\ref{eq:rhoL-alpha-beta}) can equivalently be written as
\begin{equation}\label{eq:rhoL-gamma-beta}
\widehat{\rho}_L (t) = e^{-\widehat{H}_S(\lambda_n) \beta_S(t) -\widehat{B} \gamma(t)}/\mathcal{Z}(t),
\end{equation}
where $\gamma(t)=\alpha(t) -\lambda_{t} \beta_S(t)$ was introduced earlier in Eq.~(\ref{eq:gamma-def}). While the local densities in Eqs.~(\ref{eq:rhoL-alpha-beta}) and (\ref{eq:rhoL-gamma-beta}) are equivalent, we can regard the density in these equations as expressing the behavior in different coordinate frames: in $(\alpha,\lambda)$ coordinates in Eq.~(\ref{eq:rhoL-alpha-beta}) the equilibrium states occur along the line $\alpha= \lambda \beta_S$, while in Eq.~(\ref{eq:rhoL-gamma-beta}) the equilibrium state is fixed at $\gamma=0$ since the $(\gamma, \lambda)$ coordinates correspond to a frame moving with $\lambda_t$.

One can see that the equilibrium distribution will be obtained if $\gamma(t)$ tends to zero in the time interval $\tau$. We can determine the time evolution of $\gamma(t)$ in any time interval $n$ from the evolution equations for $U(t)$ and $b(t)$. Since $\dot{U}(t)-\langle \widehat{w}_t \rangle_t=0$ for an isolated system we need only the equation of motion for $b(t)$:
\begin{eqnarray}\label{eq:b-aid}
\dot{b}(t)= {L}_{BB}(t) \gamma(t),
\end{eqnarray}
In writing this equation we used the fact that $\langle \widehat{{\mathcal J}}_B \rangle_t=0$ for an isolated system, which follows from the relation $\langle \bm{\mathcal{J}}_{s} \rangle_t \cdot \bm{\phi}_s=0$ in Eq.~(\ref{eq:avJaphi=0}) specialized to an isolated system, along with $\langle \widehat{\mathcal{J}}_U \rangle_t =0$.

Within each $\tau$ time interval where $\lambda_t$ is constant, we show in Appendix~\ref{app:gamma-decay} that $\gamma(t)$ satisfies
\begin{equation}\label{eq:gamma-mark}
\dot{\gamma}(t)= -\Gamma_{BB}(t) \gamma(t),
\end{equation}
with $\Gamma_{BB}(t)= {L}_{BB}(t)(\mathcal{K}_{SS,t}/|\bm{\mathcal{K}}_t |)$. Since $\Gamma_{BB}(t) \ge {\rm const} >0$, then $\gamma(t)$ will decay to zero. For instance, the solution of this equation in the $n^{th}$ time interval is $\gamma((n+1)\tau)=\gamma(n \tau) \exp{(-\int_{n\tau}^{(n+1)\tau} dt_1 \; \Gamma_{BB}(t_1)})$. It follows that $\gamma(t) \to 0$ as $\tau$ increases.

Also, since $\gamma(t) \to 0$ in this limit, from Eq.~(\ref{eq:sigma-adiab-b}) we have $\sigma_\gamma=0$ and the entropy production vanishes for a reversible process in an isolated system as expected.

\section{Reservoirs in the absence of coupling to the subsystem}\label{sec:reservoir}
In order to derive equations that describe the dynamics in the interfacial coupling region that separates the quantum subsystem from the quantum reservoir, it is necessary to first fully characterize the reservoir when it is decoupled from the subsystem. If the quantum subsystem is coupled to several independent reservoirs labeled by an index $k$, each reservoir $k$, when isolated from the subsystem, can be described by a Hamiltonian $\widehat{H}_k$. The reservoir density operator $\widehat{\rho}^+_k(t)$ satisfies the Liouville equation,
\begin{equation}
\partial_t \widehat{\rho}^+_k(t)=-\frac{i}{\hbar}[\widehat{H}_k,\widehat{\rho}^+_k(t)].
\end{equation}
The local density operator for the reservoir is
\begin{equation}
\widehat{\rho}^+_{Lk}(t)=e^{-\widehat{H}_k(\bm{r}) \ast \beta^+_k(\bm{r},t)}/\mathcal{Z}^+(t).
\end{equation}
The average local energy of reservoir $k$ is $e^+_k(\bm{r},t)={\rm Tr} \widehat{H}_k (\bm{r}) \widehat{\rho}^+_k(t) = {\rm Tr} \widehat{H}_k(\bm{r}) \widehat{\rho}^+_{Lk}(t)$, and satisfies the constraint condition so that its exact value is given by its local density average. We use the superscript $+$ symbol to signify that the reservoir is not interacting with the system.

An application of the formalism presented earlier, using the Markovian approximation for the dissipative kernels, shows that $e^+_k(\mathbf{r},t)$ satisfies the equation,
\begin{align}
\label{eq:k-phik}
&\dot{e}^+_k(\bm{r},t)=  -\partial_{\mathbf r} \cdot \langle \widehat{\bm{j}}^+_k(\mathbf{r}) \rangle_t   \\
& \qquad -\partial_{\bm{r}} \cdot \bm{\mathsf{L}}^+_k(\bm{r},\bm{r},_1 t) \ast \partial_{\bm{r}_1}\beta_k^+(\bm{r}_1,t)),\nonumber
\end{align}
where
\begin{equation}\label{eq:Lk+r1}
\bm{\mathsf{L}}^+_k(\bm{r},\bm{r}_1, t)=\int_0^\infty dt_1 \; \big\langle  \widehat{\mathbf{j}}^+_{k,t}(\bm{r},t_1)\overline{\mathbf{j}}^+_{k,t}(\bm{r}_1) \big\rangle^{\rm sym}_{t}.
\end{equation}
Since the energy of an isolated reservoir is conserved, we have used the flux expression, $\widehat{\mathcal{J}}^+_k(\mathbf{r})=-\partial_{\mathbf r} \cdot \widehat{\bm{j}}^+_k(\mathbf{r})$, in writing the above equations. As discussed earlier, the flux-flux correlation function will be vanishingly small for distances $|\bm{r}-\bm{r}_1|$ greater than a small microscopic length so that it may be approximated by its localized form, $\langle  \widehat{\mathbf{j}}^+_{k,t}(\bm{r},t_1)\overline{\mathbf{j}}^+_{k,t}(\bm{r}_1) \rangle^{\rm sym}_{t} \ast \partial_{\bm{r}_1}\beta_k^+(\bm{r}_1,t) \approx \langle  \widehat{\mathbf{j}}^+_{k,t}(\bm{r},t_1)\overline{\mathbf{j}}^+_{k,t} \rangle^{\rm sym}_{t} \cdot \partial_{\bm{r}}\beta_k^+(\bm{r},t)$. Implementing this localization approximation, the reservoir local energy satisfies,
\begin{equation}\label{eq:ek-betak-loc}
\dot{e}^+_k(\bm{r},t)=   - \partial_{\bm{r}} \cdot \bm{\mathsf{J}}_k^+(\bm{r},t),
\end{equation}
where the flux is
\begin{equation}\label{eq:Jk+}
\bm{\mathsf{J}}_k^+(\bm{r},t)= \bm{\mathsf{L}}^+_k(\bm{r}, t) \cdot \partial_{\bm{r}}\beta^+(\bm{r},t) ,
\end{equation}
with
\begin{equation}\label{eq:Lk+}
\bm{\mathsf{L}}^+_k(\bm{r}, t)=\int_0^\infty dt_1 \; \big\langle  \widehat{\mathbf{j}}^+_{k,t}(\bm{r},t_1)\overline{\mathbf{j}}^+_{k,t} \big\rangle^{\rm sym}_{t}.
\end{equation}
For the system studied here where there are no fluid velocity fields, the average flux $\langle \widehat{\bm{j}}^+_k(\bm{r}) \rangle_t=0$.

We can cast Eq.~(\ref{eq:ek-betak-loc}) into the form of the (generalized) heat equation.  We introduce the local temperature field by writing $\beta^+_k(\bm{r},t)=1/(k_B T_k^+(\bm{r},t))$, and the local thermal conductivity through the relation~\cite{GM84} $\mathbf{L}^+_k(\bm{r}, t)=k_B \lambda^+_k(\bm{r}, t)(T_k^+)^{2}(\bm{r},t)\bm{1}$ in the expression for the $\bm{\mathsf{J}}_k^+$ flux. We can relate the time derivative of the energy field to the local temperature by applying Eq.~(\ref{eq:adot-phidot}) to the reservoir variables to give
\begin{eqnarray}\label{eq:ek-dot-betak-dot}
\dot{e}^+_k(\bm{r},t)&=&-\langle \widetilde{H}_k(\bm{r}) \overline{H}_k(\bm{r}_1)\rangle_t \ast \dot{\beta}^+_k(\bm{r}_1,t)\nonumber \\
&=& c^+_k(\bm{r},t) \dot{T}_k^+(\bm{r},t),
\end{eqnarray}
where the generalized specific heat at constant volume is defined by $c^+_k(\bm{r},t) = \langle \widetilde{H}_k(\bm{r}) \overline{H}_k(\bm{r})\rangle_t /(k_B (T_k^+)^{2}(\bm{r},t))$.
We find that local temperature satisfies
\begin{equation}
\label{eq:k-temp-loc}
c_k^+(\bm{r},t)\dot{T}^+_k(\bm{r},t)= \partial_{\mathbf r} \lambda^+_k(\bm{r}, t) \cdot \partial_{{\mathbf r}}T^+_k(\bm{r},t) .
\end{equation}
In many applications the spatial and temporal dependence of the thermal conductivity and specific heat can be neglected, as is the case for the standard heat equation.

\section{Boundary conditions and coupled subsystem and reservoir equations}\label{sec:BC}
The interactions between the reservoirs and quantum subsystem affect the dynamics and properties of both subsystems. Even though the reservoirs are assumed to be very large, their properties in and near the interaction zones are substantially modified. The interfacial zones have lengths that are small on the scale of the reservoir dimensions. In such a circumstance their effects on the reservoirs can be replaced by boundary conditions on the reservoir equations. We derive  these boundary conditions in this section and construct the coupled equations that describe the dynamics of the quantum subsystem and reservoirs.

\subsection{Decomposition of the full system into subsystem and reservoir contributions}
It is sufficient to consider a single reservoir for this derivation, so we omit the subscript $k$ in what follows. Since the reservoirs are assumed to be independent of each other and are coupled only through their interactions with the subsystem, the development presented here can be extended to many reservoirs simply by restoring the $k$ index.

The spatial domains occupied by the quantum subsystem and reservoir can be specified through a function $h(\bm{r})$ with the properties $h(\bm{r})<0$ and $h(\bm{r})>0$ for $\bm{r}$ values in the subsystem and reservoir, respectively.~\cite{Bedeaux1976} The interface between the two systems is then described by the coordinates $\bm{r}_c$ that satisfy the condition $h(\bm{r}_c)=0$. Using this specification, we introduce a characteristic function $\theta_s(\bm{r})= H(-h(\bm{r}))$ for the subsystem that is unity for $\bm{r}$ values in quantum subsystem and zero otherwise, where $H(x)$ is the Heaviside function. The reservoir to which it is coupled may then be specified by a characteristic function $\theta_r(\bm{r})= H(h(\bm{r}))= 1- \theta_s(\bm{r})$. Far from the interaction zone with the system, this large reservoir has local energy $e^+(\bm{r},t)$ with local inverse temperature $\beta^+(\bm{r},t)$, governed by the equations discussed in the previous section.

To simplify the presentation we adopt a local Cartesian frame where $\hat{\bm{z}}$ is a local normal to the interfacial zone at $z_c$ that separates quantum subsystem from the reservoir, and $\bm{r}_\parallel$ is parallel to the interface. These considerations prompt the definitions of the following fields:
\begin{eqnarray}\label{eq:e-beta-comp}
e_{c}(\bm{r},t) &=& e_{0}(\bm{r},t)- H(z-z_c) e^+(\bm{r},t),\\
\beta_{c}(\bm{r},t) &=& \beta_{0}(\bm{r},t) - H(z-z_c) \beta^+(\bm{r},t), \nonumber
\end{eqnarray}
which, by construction, are localized to the vicinity of the coupling region. We start by writing Eq.~(\ref{eq:e0}) for $e_0(\bm{r},t)$  after making the Markovian and localization approximations on the correlation functions,
\begin{eqnarray}\label{eq:e0-ex-Mark}
\dot{e}_0(\bm{r},t)=  \langle \widehat{{J}}_0(\bm{r}) \rangle_t  &-&L_{cc}(\bm{r},t)\big( \beta_S(t) - \beta_0(\bm{r},t)\big) \nonumber \\
&-& \partial_{\bm{r}} \cdot \bm{\mathsf{L}}_{00}(\bm{r}, t)\cdot \partial_{\bm{r}} \beta_0(\bm{r},t).
\end{eqnarray}
The dissipative coefficient $L_{cc}(\bm{r},t)$ is defined in a manner similar to that in Eq.~(\ref{eq:Lcc-coef-r}) except the first flux has $\bm{r}$ dependence,
\begin{equation}
{L}_{cc}(\bm{r},t)=\int_0^\infty dt_1 \;\langle  \widehat{J}_{c,t}(\bm{r},t_1,t) \overline{J}_{c,t} \rangle^{\rm sym}_{t},
\end{equation}
while the $\bm{\mathsf{L}}_{00}$ flux correlation function was defined previously in Eq.~(\ref{eq:L00}).

Inserting the decomposition of the fields given in Eqs.~(\ref{eq:e-beta-comp}) into Eq.~(\ref{eq:e0-ex-Mark}), we obtain,
\begin{eqnarray} \label{eq:ec}
\dot{e}_{c}(\bm{r},t)&=& \langle \widehat{{J}}_0(\bm{r}) \rangle_t  +\mathsf{J}^{(0)}_{c}(\bm{r},t) -\partial_{\bm{r}} \cdot \bm{\mathsf{J}}^{(1)}_{c}(\bm{r},t)\nonumber  \\
&&- \delta(z-z_c) \hat{\bm{z}} \cdot \bm{\mathsf{J}}^+(\bm{r},t) .
\end{eqnarray}
By construction, all of the $\bm{r}$ coordinates in Eq.~(\ref{eq:ec}) are confined to lie in the vicinity of the interaction zone.

The $\mathsf{J}^{(0)}_{c}(\bm{r},t)$ and $\bm{\mathsf{J}}^{(1)}_{c}(\bm{r},t)$ fluxes are
\begin{eqnarray}\label{eq:J0c-4}
&&\mathsf{J}^{(0)}_{c}(\bm{r},t)=  -{L}_{cc}(\bm{r},t) \big(\beta_S(t) - \beta_c(t)\big), \nonumber\\
&&\bm{\mathsf{J}}^{(1)}_{c}(\bm{r},t)= \bm{\mathsf{L}}_{00}(\bm{r}, t) \cdot \partial_{\bm{r}} \beta_c(\bm{r},t),
\end{eqnarray}
and we note that the $\bm{\mathsf{J}}^{(1)}_{c}(\bm{r},t)$ flux has a form analogous to $\bm{\mathsf{J}}^+(\bm{r},t)$ that was defined in Eq.~(\ref{eq:Jk+}). Additional details of the steps leading to this result are given in Appendix~\ref{app:gen-BC}.

Integrating Eq.~(\ref{eq:ec}) over the interfacial region, we obtain
\begin{eqnarray} \label{eq:ec-int}
\dot{e}_{c}(\bm{r}_\parallel,t)&=& \langle \widehat{{J}}_0(\bm{r}_\parallel) \rangle_t  +\mathsf{J}^{(0)}_{c}(\bm{r}_\parallel,t) -\partial_{\bm{r}_\parallel} \cdot \bm{\mathsf{J}}^{(1)}_{c}(\bm{r}_\parallel,t)\nonumber  \\
&&- \hat{\bm{z}} \cdot \bm{\mathsf{J}}^+(z_c,\bm{r}_\parallel,t) .
\end{eqnarray}
We can write Eq.~(\ref{eq:ec-int}) as
\begin{eqnarray} \label{eq:BC0}
\hat{\bm{z}} \cdot \bm{\mathsf{J}}^+(z_c,\bm{r}_\parallel,t)&=& \mathsf{J}^{(0)}_{c}(\bm{r}_\parallel,t) -\Sigma_c(\bm{r}_\parallel,t),
\end{eqnarray}
where $\Sigma_c$ represents the heat equation in the interaction zone,
\begin{equation}\label{eq:Sigma_C}
\Sigma_c(\bm{r}_\parallel,t)=\dot{e}_{c}(\bm{r}_\parallel,t) - \langle \widehat{{J}}_0(\bm{r}_\parallel) \rangle_t +\partial_{\bm{r}_\parallel} \cdot \bm{\mathsf{J}}^{(1)}_{c}(\bm{r}_\parallel,t).
\end{equation}
The boundary condition on the reservoir energy field follows by setting $\Sigma_c =0$ to give,
\begin{equation} \label{eq:BC-res}
 \hat{\bm{z}} \cdot \bm{\mathsf{J}}^+(z_c,\bm{r}_\parallel,t)= {L}_{cc}(\bm{r}_\parallel,t) \big( \beta_S(t) -  \beta_c(t)\big).
\end{equation}
This boundary condition accounts for the coupling between the reservoir temperature field and the quantum subsystem. As mentioned earlier, in many applications the equations can be simplified by neglecting the spatial dependence of the transport properties. The reservoir equations (\ref{eq:ek-betak-loc}) or equivalently (\ref{eq:k-temp-loc}) may then be solved subject to this boundary condition to obtain the nonlocal reservoir temperature fields.

\section{Heat flux through a small system coupled to two large reservoirs}
\begin{figure}[htbp]
\centering
\resizebox{0.9\columnwidth}{!}{
      \includegraphics{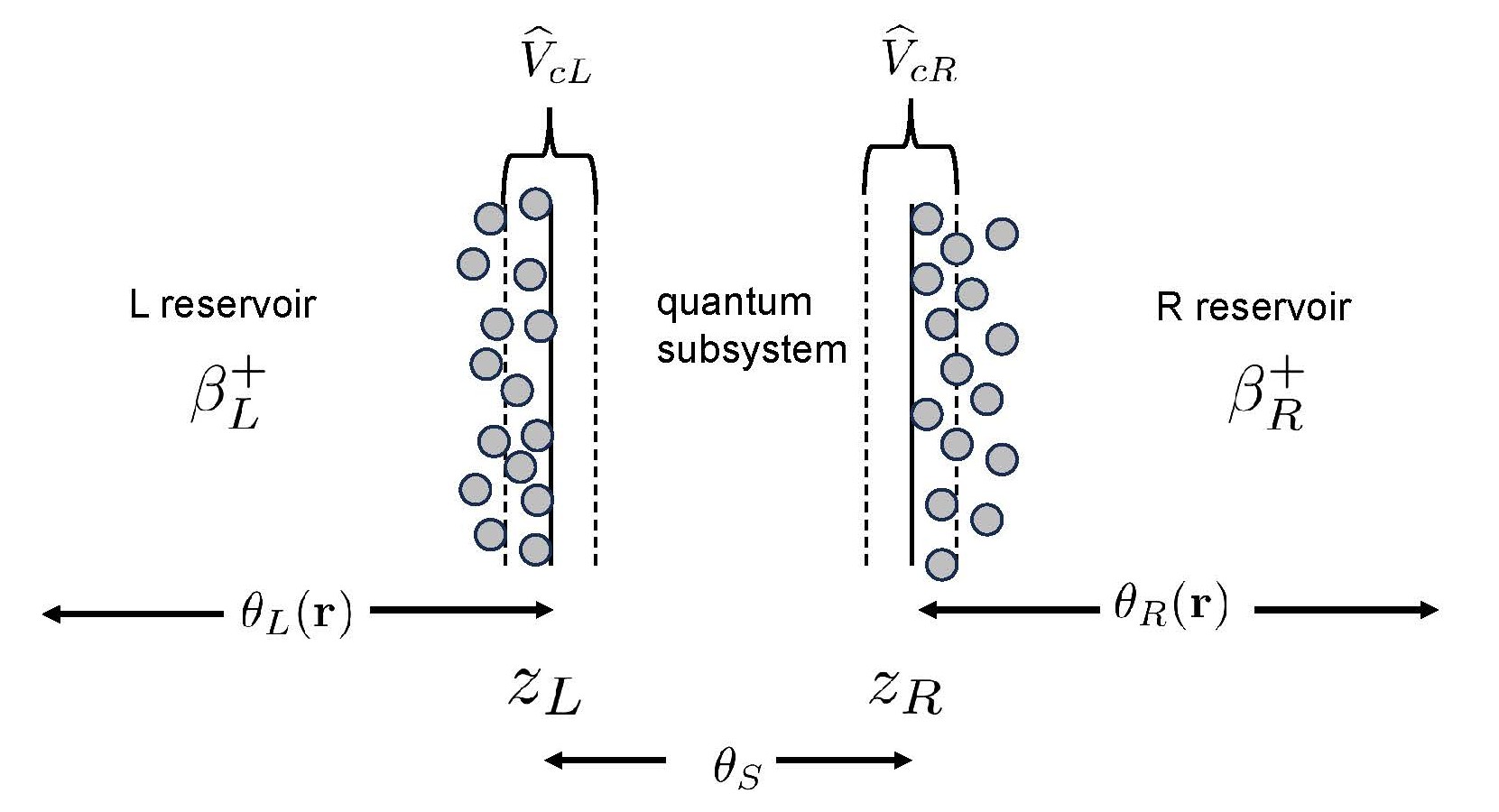}}
        \caption{Schematic diagram showing the quantum subsystem and reservoir particles (grey filled circles) near the interaction zones. The subsystem is confined to the spatial domain lying between $z_L$ and $z_R$, while the reservoirs lie outside this region. The coupling regions where $\widehat{V}_{cL}$ and $\widehat{V}_{cR}$ are nonzero are indicated by vertical dashed lines. The inverse temperatures of the reservoirs far from the interaction regions are denoted by $\beta^+_{R,L}=1/k_B T^+_{R,L}$.
    } \label{fig:two-baths-schm-S}
\end{figure}

The heat flux through molecular junctions has been studied extensively, including detailed investigations of molecular aspects of the mechanisms that underlie the heat transport in such systems, as documented and  summarized in reviews.~\cite{heat-transport-ARPChem-2016,heat-transport-RMP-2021} Here, to illustrate the application of the general formalism described in the previous section, we examine gross aspects of heat flux through a small quantum system coupled to two large reservoirs with different temperatures, as sketched in Fig.~\ref{fig:two-baths-schm-S}.
The Hamiltonian for the entire system is
\begin{eqnarray}
\widehat{\mathsf{H}}&=& \widehat{H}_S + \widehat{V}_{L} + \widehat{H}_L + \widehat{V}_{R} +\widehat{H}_R \nonumber \\
&\equiv& \widehat{H}_S + \widehat{H}_{0L} +\widehat{H}_{0R}.
\end{eqnarray}
In this application we do not include external driving so no work is performed. We assume that the two quantum reservoirs are large, with left and right temperatures $T^+_{L,R}=1/k_B\beta_{L,R}^+$ far from the interaction region with the quantum system. The small quantum system, which is confined to lie between $z_L$ and $z_R$, interacts with these reservoirs through the $\widehat{V}_{L,R}$ coupling operators that have short range. The geometry is as shown in the figure so that the characteristic functions $\theta_{L,R}(z)$ depend only on the $z$ coordinate and take the forms, $\theta_{L}(z)=H(z_L-z)$ and $\theta_{R}(z)=H(z-z_R)$, where $H(z)$ is again the Heaviside function.

To carry out the analysis of the heat flow in this system, we make use of the boundary condition equations derived in the previous section. From Eq.~(\ref{eq:BC-res}) the boundary conditions on the reservoir equations are:
\begin{equation} \label{eq:BC-res2}
\hat{\bm{z}}\cdot \bm{\mathsf{J}}_k^+(z_{k},t)= {L}_{cc,k}(t) \big( \beta_S(t) -  \beta_{ck}(t)\big),
\end{equation}
with $ \hat{\bm{z}}$ a unit vector. The flux is given by
\begin{eqnarray}
\mathsf{J}_k^+(z_{k},t)&=& \mathsf{L}^+_k(z, t) \partial_z\beta_k^+(z,t) |_{z_{k}} \\
&=& - \lambda^+_k(z, t) \partial_z T^+_k(z,t) |_{z_{k}}\nonumber .
\end{eqnarray}
Approximating transport coefficients and specific heats by their constant values, the two reservoirs are described by the equations of motion,
\begin{eqnarray}
c_k^+\partial_t T^+_k(z,t)=  \lambda^+_k \partial^2_z T^+_k(z,t),
\end{eqnarray}
 with their boundary conditions,
\begin{eqnarray}
\partial_z T^+_k(z_k,t) &=&\pm ({L}_{cc,k}/\lambda^+_k k_B) \big( 1/T_S(t) -  1/T_{ck}(t)\big)\nonumber\\
T^+_k(\it{Z}_k,t)&=& T^+_k,
\end{eqnarray}
where the minus sign is for $k=L$ and the plus sign for $k=R$, and $\it{Z}_k$ is the outer boundary of reservoir $k$ where the temperature takes the value $T^+_k$.

\subsection{Steady state heat flux}

The time evolution of the internal energy is obtained by writing Eq.~(\ref{eq:U2-approx}) for this system,
\begin{eqnarray}\label{eq:U2-2res}
\partial_t U(t) = \sum_{k=L}^R\big[- \langle \widehat{J}_{ck} \rangle_t + {L}_{cc,k}(t)\big(\beta_S(t) -\beta_{ck}(t)\big)\Big].\nonumber \\
\end{eqnarray}
The equation of motion for $e_c(t)$ can be obtained by integrating evolution equation for $e_c(t)(\bm{r}_\parallel,t)$ over $\bm{r}_\parallel$ to give $\dot{e}_{c}(t) = \langle \widehat{J}_{ck} \rangle_t$. From these equations, in the steady state we have
\begin{eqnarray}
 \xi \equiv {L}_{cc,L}\big(\beta_S -\beta_{cL}\big) ={L}_{cc,R}\big(\beta_{cR}-\beta_S\big).\nonumber
\end{eqnarray}
We suppose that $T_L^+ > T_R^+$ and compute the steady state heat flux through the system. Since $\frac{d^2}{dz^2} T^+_k(z)=0$ in the steady state, we have $\frac{d}{dz} T^+_k(z) ={\rm const}$ so that $T^+_k(z)$ is a linear function of $z$. Using the boundary conditions we get
\begin{eqnarray}
T_L^+(z)&=& T_L^+ - \frac{\xi}{\lambda^+_L } (z-\it{Z}_L)\\
T_R^+(z)&=& T_R^+ - \frac{\xi}{\lambda^+_R } (z-\it{Z}_R).
\end{eqnarray}
Since the temperatures must be positive, we have $T_{cL}>T_S>T_{cR}$. Also recall that since we have taken the origin of the coordinate system at the center of the $S$ region, $z$ (and $\it{Z}_L=-\it{Z}_R)$ are negative in the $L$ reservoir.

The heat flux through the system is controlled by the flux at the boundaries of the reservoirs with the quantum subsystem, and using the relations given above we have
\begin{equation}
J_L^+(z_L)={L}_{cc,L} \big( \beta_S -  \beta_{cL}\big)={L}_{cc,R} \big(\beta_{cR}- \beta_S \big)=J_R^+(z_R).
\end{equation}
The value of $(\beta_S -  \beta_{cL,R})$ can be obtained from the solution of Eq.~(\ref{eq:DeltaSc}) given in Appendix~\ref{app:constraint-var}.

\section{Conclusion}
The derivation of quantum thermodynamics of a small system open to nonequilibrium quantum reservoirs was carried out using a set of quantum observables whose averages are directly related to the thermodynamic quantities of interest: internal energy and entropy. The derivation yielded generalized forms of the thermodynamic laws. The derivation made use of a local density operator constructed so as to give the exact nonequilibrium averages of the selected set of operators. The resulting equations are  nonlinear and nonlocal generalizations of the phenomenological evolution equations for the conserved or nearly conserved variables. Expressions for the work and heat based on these generalized equations were derived and analyzed.

The latter sections of the paper showed how the coupling of large quantum reservoirs to the subsystem could be described through boundary conditions on the reservoir fields in the absence of the quantum subsystem. The boundary conditions involve the constraint variables of the subsystem and interaction regions and contain dissipative coefficients whose values are specified by flux-flux correlation functions in the interaction regions.

In this work, the operators in set of variables were chosen to be appropriate for a small quantum system coupled to large reservoirs, in particular, the quantum subsystem was described by its global Hamiltonian $\widehat{H}_S$ while the reservoirs with coupling were described by the local Hamiltonian density $\widehat{H}_0(\bm{r})$. These restrictions are easily relaxed. If the quantum subsystem is large enough to merit a local description the subsystem Hamiltonian density $\widehat{H}_S(\bm{r})$ can be used, and if the reservoirs are fluids additional conserved densities can be included in the set.

The focus in this paper was on the general formulation of quantum thermodynamics in terms of the local density operators. Although the formalism is constructed to only give the exact average values of the selected set of operators, once the relation between the full and local density operators is established, the equations of motion for other variables can be constructed, although their average values are no longer given exactly by the local density operator. This allows one to investigate other aspects of the quantum dynamics of subsystems coupled to nonequilibrium reservoirs, such as population and coherence evolution and entanglement between the system and reservoir. In addition, other detailed aspects of quantum thermodynamics that have been investigated by other means could be studied in the context of the formalism used here.

\section*{Acknowledgements}
Financial support was received from the Natural Sciences and Engineering Research Council of Canada and the Leverhulme Trust, RPG-2023-078.

\appendix

\section{Projection and density operators}\label{app:projectors}
The projection operator $\widehat{\mathcal{P}}^\dagger(t)$ defined in Eq.~(\ref{eq:Pdag-proj}) maps the full density operator onto the local density operator, yielding
\begin{align}
\label{eq:density-split}
\widehat{\rho}(t)=\widehat{\mathcal{P}}^\dagger(t)\widehat{\rho}(t)
+\widehat{\mathcal{Q}}^\dagger(t)\widehat{\rho}(t) =\widehat{\rho}_L(t)+\widehat{\mathcal{Q}}^\dagger(t)\widehat{\rho}(t).
\end{align}
Thus, $\widehat{\mathcal{Q}}^\dagger(t)\widehat{\rho}(t)$ must be evaluated to determine the relation between the full and local densities. For this purpose its equation of motion is constructed and formally solved. The time derivative of $\widehat{\mathcal{Q}}^\dagger(t)\widehat{\rho}(t)$ can be written as
\begin{align}
\label{eq:timeder-Qrho}
\partial_t \widehat{\mathcal{Q}}^\dagger(t)\widehat{\rho}(t)=\partial_t \widehat{\rho}(t) - \partial_t ({\mathcal P}^\dagger (t) \widehat{\rho}(t)).
\end{align}
We can then make use of the following properties of the projector: ${\mathcal P}^\dagger(t) \partial_t \widehat{\rho}_L(t) = \partial_t \widehat{\rho}_L(t)$, which can be confirmed by applying the definition of the projection operator in Eq.~(\ref{eq:Pdag-proj}) to the time derivative of $\widehat{\rho}_L(t)$ in Eq.~(\ref{eq:drhoL-dt}). Since ${\rm{Tr}} \big( \partial_t \widehat{\rho}(t) \bm{\widehat{A}}(\bm{r}) \big) = {\rm{Tr}} \big( \partial_t \widehat{\rho}_L(t) \bm{\widehat{A}}(\bm{r}) \big)$ because of the constraint condition, we also have ${\mathcal P}^\dagger (t) \partial_t \widehat{\rho}(t) = \partial_t \widehat{\rho}_L (t)
= \partial_t \, {\mathcal P}^\dagger (t) \widehat{\rho}(t)$. These relations allow us to write Eq.~(\ref{eq:timeder-Qrho}) as
\begin{eqnarray}
\partial_t \widehat{\mathcal{Q}}^\dagger(t) \widehat{\rho}(t)
    &=& -\widehat{\mathcal{Q}}^\dagger(t) i\widehat{\mathcal{L}}(\lambda_t) \widehat{\rho}_L(t)\nonumber \\
&&- \widehat{\mathcal{Q}}^\dagger(t) i \widehat{\mathcal{L}}(\lambda_t) \widehat{\mathcal{Q}}^\dagger(t) \widehat{\rho}(t),
\end{eqnarray}
whose formal solution is
\begin{align}
\widehat{\mathcal{Q}}^\dagger(t) \widehat{\rho}(t)
    &= \widehat{\mathcal{U}}^\dagger_Q(t,0)\widehat{\mathcal{Q}}^\dagger(0) \widehat{\rho}(0) \nonumber \\
    &\quad - \int_0^t dt_1 \;
       \widehat{\mathcal{U}}^\dagger_Q(t,t_1) \widehat{\mathcal{Q}}^\dagger(t_1) i\widehat{\mathcal{L}}(\lambda_{t_1}) \widehat{\rho}_L(t_1).
\end{align}
Substitution into Eq.~(\ref{eq:density-split}) yields the result given in Eq.~(\ref{eq:rho-rhoL}) in the main text.

\section{Evolution equations for average values} \label{app:av-value-eqs}
Additional details of derivation leading to the equations of motion for the average observables in Eq.~(\ref{eq:Qa-phi}) are given here. When Eq.~(\ref{eq:rho-rhoL}) for $\widehat{\rho}(t)$ is inserted in Eq.~(\ref{eq:dadt}) for the time derivative of the average observable, we find
\begin{eqnarray}
&&\dot{\bm{a}}(\bm{r},t)= \langle \widehat{\bm{\mathcal{J}}}_{A}(\bm{r}) \rangle_t + {\rm Tr} \big( \widehat{\bm{\mathcal{J}}}_{A}(\bm{r}) \widehat{\mathcal{\mathcal{U}}}^\dagger_Q(t,0)\widehat{\mathcal{Q}}^\dagger(0) \widehat{\rho}(0) \big) \nonumber \\
&&\quad -\int_0^t dt_1 \; {\rm Tr} \big( \widehat{\bm{\mathcal{J}}}_{A}(\bm{r})\widehat{\mathcal{U}}^\dagger_Q(t,t_1) \widehat{\mathcal{Q}}^\dagger(t_1) i\widehat{\mathcal{L}}(t_1) \widehat{\rho}_L(t_1)\big), \nonumber
\end{eqnarray}
This equation can be written a more convenient form by moving the evolution operator onto the first flux to obtain,
\begin{eqnarray}
&&{\rm Tr} \big( \widehat{\bm{\mathcal{J}}}_{A}(\bm{r})\widehat{\mathcal{U}}^\dagger_Q(t,t_1) \widehat{\mathcal{Q}}^\dagger(t_1) i\widehat{\mathcal{L}} \widehat{\rho}_L(t_1)\big)=\\
&& \qquad \qquad {\rm Tr} \big( \widehat{\bm{\mathcal J}}_{A,t}(\bm{r},t_1,t) \widehat{\mathcal{Q}}^\dagger(t_1) i\widehat{\mathcal{L}} \widehat{\rho}_L(t_1)\big), \nonumber
\end{eqnarray}
where $\widehat{\bm{{\mathcal J}}}_{A,t}(\bm{r},t_1,t) =  \widehat{\mathcal{U}}_{Q}(t_1,t) \widehat{\bm{\mathcal{J}}}_{A,t}(\bm{r})$ and $\widehat{\bm{\mathcal{J}}}_{A,t}(\bm{r}) \equiv \widehat{\mathcal{Q}}(t)\widehat{\bm{\mathcal{J}}}_{A}(\bm{r})$ previously defined in the main text. The projected evolution operator is
\begin{equation}
\widehat{\mathcal{U}}_Q(t,t_1)=\mathcal{T}_- \exp \Big(\int_{t_1}^t dt' \;i\widehat{\mathcal{L}} (\lambda_{t'}) \widehat{\mathcal{Q}}(t') \Big),
\end{equation}
and the time ordering operator $\mathcal{T}_-$ signifies that larger time arguments appear to the right of smaller time arguments. An expression for $\widehat{\mathcal{Q}}^\dagger(t_1) i\widehat{\mathcal{L}}(t_1) \widehat{\rho}_L(t_1)$ is needed to complete the derivation. We first compute $i\widehat{\mathcal{L}}(\lambda_t) \widehat{\rho}_L(t)$ to obtain,
\begin{eqnarray}
&&i\widehat{\mathcal{L}}(\lambda_t) \widehat{\rho}_L(t) = \frac{1}{\mathcal{Z}(t)}
\frac{i}{\hbar} [\widehat{\mathsf{H}}(\lambda_t), e^{\widehat{\bm{A}}(\bm{r}) \ast \bm{\phi}(\bm{r},t)}]\\
&& \quad = \frac{1}{2}\Big(\overline{\bm{\mathcal{J}}}_{A} (\bm{r})  \widehat{\rho}_L(t) +\widehat{\rho}_L(t) \overline{\bm{\mathcal{J}}}^\dagger_{A} (\bm{r})\Big) \ast \bm{\phi} (\bm{r},t) \nonumber
\end{eqnarray}
by making use of the identity
\begin{equation}\label{eq:commutator-relation}
[\widehat{\mathsf{H}}(\lambda_t), e^{\widehat{\bm{A}} \ast \bm{\phi}}]= \overline{[\widehat{\mathsf{H}}(\lambda_t),\widehat{\bm{A}}]} \ast \bm{\phi}\; e^{\widehat{\bm{A}} \ast \bm{\phi}}.
\end{equation}

The last step is to evaluate the action of $\widehat{\mathcal{Q}}^\dagger(t_1)$ on this quantity. This is accomplished with the aid of the following projection operator identities:
\begin{eqnarray}
&&\widehat{\mathcal{Q}}^\dagger(t) \overline{\bm{\mathcal{J}}}_A(\bm{r}) \widehat{\rho}_L(t)  =
\big( \overline{\mathcal{Q}}(t)\overline{\bm{\mathcal{J}}}_{A}\big) \widehat{\rho}_L(t) \equiv  \overline{\bm{\mathcal{J}}}_{A,t} \widehat{\rho}_L(t),\nonumber \\
&&  \widehat{\mathcal{Q}}^\dagger(t)\Big( \widehat{\rho}_L(t) \overline{\bm{\mathcal{J}}}^\dagger_{A}\Big)=  \widehat{\rho}_L(t) \big(\overline{Q}^\dagger(t)\overline{\bm{\mathcal{J}}}^\dagger_{A} \big) \equiv \widehat{\rho}_L(t)  \overline{\bm{\mathcal{J}}}^\dagger_{A,t} ,\nonumber \\
\end{eqnarray}
where the projection operators $\overline{Q}(t) = 1 - \overline{P}(t)$ and $\overline{Q}^\dagger (t) = 1 - \overline{P}^\dagger (t)$ are defined by
\begin{align}
\overline{P}(t)\overline{\bm{\mathcal{J}}}_{A} & =
\langle \overline{\bm{\mathcal{J}}}_{A} \rangle_t + \overline{\bm{A}} \ast \langle
\widetilde{\bm{A}}\overline{\bm{A}}\rangle_t^{-1}
 \ast \langle \widetilde{\bm{A}} \overline{\bm{\mathcal{J}}}_{A}  \rangle_t \\
\overline{P}^\dagger (t)\overline{\bm{\mathcal{J}}}^\dagger_{A} &=
\langle \overline{\bm{\mathcal{J}}}^\dagger_{A} \rangle_t +  \langle  \overline{\bm{\mathcal{J}}}^\dagger_{A} \widetilde{\bm{A}}  \rangle_t \ast \langle \widetilde{\bm{A}} \overline{\bm{A}} \rangle_t^{-1}  \ast \overline{\bm{A}}^\dagger.
\end{align}
The resulting expression is
\begin{eqnarray}
&&\widehat{\mathcal{Q}}^\dagger (t) i\widehat{\mathcal{L}}(\lambda_t) \widehat{\rho}_L(t)= \\
&& \qquad \frac{1}{2}\Big(\overline{\bm{\mathcal{J}}}_{A,t}(\bm{r})  \widehat{\rho}_L(t) +\widehat{\rho}_L(t) \overline{\bm{\mathcal{J}}}^\dagger_{A,t} (\bm{r})\Big) \ast \bm{\phi} (\bm{r},t),\nonumber
\end{eqnarray}
whose use leads to Eq.~(\ref{eq:Qa-phi}) that contains the symmetrized dissipative matrix defined by
\begin{eqnarray}
 &&\big\langle  \widehat{\bm{{\mathcal J}}}_{A,t}(\bm{r},t_1,t)\overline{\bm{\mathcal{J}}}_{A,t_1}(\bm{r}_1)   \big\rangle^{\rm sym}_{t_1} \equiv \\
&& \qquad \qquad
 \frac{1}{2} \Big( \big\langle  \widehat{\bm{{\mathcal J}}}_{A,t}(\bm{r},t_1,t)\overline{\bm{\mathcal{J}}}_{A,t_1}(\bm{r}_1)   \big\rangle_{t_1} \nonumber
  \\
 && \qquad \qquad+ \big\langle \overline{\bm{\mathcal{J}}}^\dagger _{A,t_1}(\bm{r}_1) \widehat{\bm{{\mathcal J}}}_{A,t}(\bm{r},t_1,t)   \big\rangle_{t_1} \Big) . \nonumber
\end{eqnarray}.

\section{Computation of $\langle \widehat{{w}}_{t} \rangle_t$ for the specific work protocol}\label{app:adiab-local}
The work can be expressed equivalently in terms of averages over the local equilibrium distribution,
\begin{equation}
W(t)=\int_0^t dt_1 \;\langle \widehat{w}_{t_1} \rangle_{t_1},
\end{equation}
and here we show that this is also equal to the change in the internal energy. For the same work protocol as in the text, we can write
\begin{eqnarray}
W(t_f)&=& \sum_{n=1}^m \int_0^{t_f} dt \; \delta(t-n\tau)(\lambda_n-\lambda_{n-1}) {\rm Tr} \big(\widehat{B} \widehat{\rho}_L(t)\big) \nonumber \\
&=&\sum_{n=1}^m {\rm Tr} \big( (\lambda_n \widehat{B} -\lambda_{n-1} \widehat{B}) \widehat{\rho}_L(n\tau)\big) \\
&=&\sum_{n=1}^m  {\rm Tr} \big( (\widehat{H}_S(\lambda_n)  -\widehat{H}_S(\lambda_{n-1})) \widehat{\rho}_L(n\tau)\big). \nonumber
\end{eqnarray}
To write the last line in this equation, we added and subtracted $\widehat{H}_S$. Consider the time segment from $t=(n-1)\tau$ to $t=n\tau_-$.  During this time interval the system evolution is governed by $\lambda_{n-1}$ and the Hamiltonian $\widehat{H}_S(\lambda_{n-1})$ is conserved. Consequently, ${\rm Tr} \widehat{H}_S(\lambda_{n-1})\widehat{\rho}_L(n\tau)= {\rm Tr} \widehat{H}_S(\lambda_{n-1})\widehat{\rho}_L((n-1)\tau) = \langle \widehat{H}_S(\lambda_{n-1}) \rangle_{n-1}$. Therefore, we can write
\begin{eqnarray}
W(t_f)&=& \sum_{n=1}^m \big(\langle \widehat{H}_S(\lambda_n) \rangle_n - \langle \widehat{H}_S(\lambda_{n-1}) \rangle_{n-1} \big)\nonumber \\
&=&U(m\tau)-U(0).
\end{eqnarray}

\section{Equation of motion for $\gamma(t)$ in an isolated system} \label{app:gamma-decay}

To obtain the equation of motion for $\gamma(t)$ we make use of Eq.~(\ref{eq:adot-phidot}) to write the following relations:
\begin{eqnarray}
\dot{b}(t) &=&
 - \langle \widehat{B} \overline{B} \rangle_t \dot{\alpha}(t) - \langle \widehat{B} \overline{H}_S \rangle_t \dot{\beta}_S(t), \\
\dot{e}_S(t) &=&  - \langle \widehat{H}_s \overline{B} \rangle_t \dot{\alpha}(t)- \langle \widehat{H}_S \overline{H}_S \rangle_t \dot{\beta}_S(t).\nonumber
\end{eqnarray}
When expressed in $(U,b)$ variables this equation reads,
\begin{eqnarray}
\dot{b}(t) &=&  - \langle \widehat{B} \overline{B} \rangle_t (\dot{\gamma}(t)+ \dot{\lambda}_t \beta_S(t))\label{eq:db-gamma-betas} \\
&& \qquad \qquad - \langle \widehat{B} \overline{H}_S(\lambda_t) \rangle_t \dot{\beta}_S(t), \nonumber  \\
 0 &=&  - \langle \widehat{H}_S(\lambda_t) \overline{B} \rangle_t (\dot{\gamma}(t)+\dot{\lambda}_t {\beta}_S(t)) \label{eq:dU-gamma-betas}\\
&& \qquad \qquad- \langle \widehat{H}_S(\lambda_t) \overline{H}_S(\lambda_t) \rangle_t \dot{\beta}_S(t), \nonumber
\end{eqnarray}
where we used the fact that $\dot{U}(t)- \langle \widehat{w}_t \rangle_t=0$ for an isolated system. From Eq.~(\ref{eq:dU-gamma-betas}) we have
\begin{equation}
 \dot{\beta}_S(t) = -\frac{\langle \widehat{H}_S(\lambda_t) \overline{B} \rangle_t}{\langle \widehat{H}_S(\lambda_t) \overline{H}_S(\lambda_t) \rangle_t} (\dot{\gamma}(t)+\dot{\lambda}_t \beta_S(t)).
\end{equation}
Inserting this result in Eq.~(\ref{eq:db-gamma-betas}) and using the result for $\dot{b}(t)$ in Eq.~(\ref{eq:b-aid}) we find
\begin{equation}\label{eq:gamma-evol}
\dot{\gamma}(t)= -\dot{\lambda}_t \beta_S(t) -\Gamma_{BB}(t) \gamma(t),
\end{equation}
with $\Gamma_{BB}(t)= \big(\mathcal{K}_{SS,t}/|\bm{\mathcal{K}}_t |\big)\mathsf{L}_{BB}(t)$, and $\bm{\mathcal{K}}_t$ is the overlap matrix defined by Eqs~(\ref{eq:db-gamma-betas}) and (\ref{eq:dU-gamma-betas}) and $|\bm{\mathcal{K}}_t |$ is its determinant.

Since $\lambda_t$ is constant within each $\tau$ interval and $\dot{\lambda}_t=0$, within these intervals, we have Eq.~(\ref{eq:gamma-mark}) in the text.

\section{Interaction zone equations and boundary conditions}\label{app:gen-BC}

The equation of motion for $\dot{e}_c$ in Eq.~(\ref{eq:ec}) and the expressions for the fluxes in this equation follow from the following calculations. We can write this equation as
\begin{eqnarray} \label{eq:ec-0}
&&\dot{e}_{c}(\bm{r},t)= \langle \widehat{{J}}_0(\bm{r}) \rangle_t  +\mathsf{J}^{(0)}_{c}(\bm{r},t) \\ && \quad -\partial_{\bm{r}} \cdot \big(\bm{\mathsf{J}}^{(1)}_{0}(\bm{r},t)-H(z-z_c) \bm{\mathsf{J}}^+(\bm{r},t) \big) \nonumber \\
&&\quad - \hat{\bm{z}} \delta(z-z_c)) \cdot \bm{\mathsf{J}}^+(\bm{r},t).\nonumber
\end{eqnarray}
The flux $\bm{\mathsf{J}}^{(1)}_{c}=\bm{\mathsf{J}}^{(1)}_{0}(\bm{r},t)-H(z-z_c) \bm{\mathsf{J}}^+(\bm{r},t)$ appears Eq.~(\ref{eq:ec}) and can be rearranged as follows:
\begin{eqnarray}
&&\bm{\mathsf{J}}^{(1)}_{c}(\bm{r},t)= \bm{\mathsf{L}}_{00}(\bm{r}, t) \cdot \partial_{\bm{r}} \beta_0(\bm{r},t)\\
&&\qquad - H(z-z_c) \bm{\mathsf{L}}^+(\bm{r}, t) \cdot \partial_{\bm{r}} \beta^+(\bm{r},t)\nonumber \\
&& = \bm{\mathsf{L}}_{00}(\bm{r}, t) \cdot \partial_{\bm{r}} \beta_c(\bm{r},t) \nonumber \\
&&  + \big(\bm{\mathsf{L}}_{00}(\bm{r}, t)H(z-z_c)-H(z-z_c)\bm{\mathsf{L}}^+(\bm{r}, t) \big)\cdot \partial_{\bm{r}} \beta^+(\bm{r},t)\nonumber \\
&&  +\bm{\mathsf{L}}_{00}(\bm{r}, t)(\partial_{\bm{r}} H(z-z_c))\beta^+(\bm{r},t) \nonumber\\
&&  \approx \bm{\mathsf{L}}_{00}(\bm{r}, t) \cdot \partial_{\bm{r}} \beta_c(\bm{r},t). \nonumber
\end{eqnarray}
The last approximate equality follows from the observation that contributions containing products of terms that are confined to the interaction region where $\widehat{V}_c(\bm{r}) \neq 0$ and terms involving $H(z-z_c)$ or its derivative that restrict the coordinates to lie in the reservoir region will be negligibly small and may be neglected.

\section{Constraint variables}\label{app:constraint-var}
To determine the constraint variables $(\gamma(t),\beta_S(t),\beta_c(\bm{r}_c,t))$ we make use of Eq.~(\ref{eq:adot-phidot}) that expresses the time derivatives of the average fields in terms of the derivatives of the constraint variables,as discussed in Sec.~\ref{subsec:constraint} in general terms.
We have
\begin{eqnarray}
\dot{b}(t)&=& -\langle \widehat{B} \overline{B} \rangle_t \dot{\alpha}(t) - \langle \widehat{B} \overline{H}_S \rangle_t \dot{\beta}_S(t) \\
\dot{e}_S(t)&=& -\langle \widehat{H}_S \overline{B} \rangle_t \dot{\alpha}(t) - \langle \widehat{H}_S \overline{H}_S \rangle_t \dot{\beta}_S(t) \\
&& \qquad - \langle \widehat{H}_S \overline{H}_0(\bm{r}) \rangle_t \ast \dot{\beta}_0(\bm{r},t)\nonumber \\
\dot{e}_0(t)&=&  - \langle \widehat{H}_0 \overline{H}_S \rangle_t \dot{\beta}_S(t) \\
&& \qquad - \langle \widehat{H}_0 \overline{H}_0(\bm{r}) \rangle_t \ast \dot{\beta}_0(\bm{r},t).\nonumber
\end{eqnarray}
While the analysis can be carried out using this full set of matrix equations it is instructive to make some simplifying assumption to present the results in a more transparent form. We retain the dominant diagonal overlap terms and follow the same procedure as in the text to construct equations for $\dot{e}_c(t)$ and $\beta_c(t)$. Then, after rearranging to construct equations for $(b(t),U(t),e_c(t)$, we obtain
\begin{eqnarray}
\dot{b}(t)&=& -\langle \widehat{B} \overline{B} \rangle_t \dot{\gamma}(t)  \\
\dot{U}(t)-\langle \widehat{w}_t \rangle_t &=&  - \langle \widehat{H}_S(\lambda_t) \overline{H}_S(\lambda_t) \rangle_t \dot{\beta}_S(t) \\
\dot{e}_c(t)&=&  - \langle \widehat{V}_c \overline{V}_c \rangle_t  \dot{\beta}_c(t).
\end{eqnarray}
Using these expressions in conjunction with the equations of motion for $b(t)$, $U(t)$ and $e_c(t)$ in Eqs.~(\ref{eq:b2}), (\ref{eq:U2-approx}) and (\ref{eq:Sigma_C}), we obtain
\begin{eqnarray}
 -\langle \widehat{B} \overline{B} \rangle_t \dot{\gamma}(t)&=&  \langle \widehat{J}_B \rangle_t  + {L}_{BB}(t) \gamma(t) \\
 - \langle \widehat{H}_S(\lambda_t) \overline{H}_S(\lambda_t) \rangle_t \dot{\beta}_S(t) &=& -\langle \widehat{J}_c \rangle_t \\
 &+& {L}_{cc}(t)\big(\beta_S(t) -\beta_c(t)\big) \nonumber \\
 - \langle \widehat{V}_c \overline{V}_c \rangle_t  \dot{\beta}_c(t)&=& \langle \widehat{J}_c \rangle_t.
\end{eqnarray}
The latter two equations can be rearranged to find an equation for $\Delta_{Sc}=\beta_S(t) -\beta_c(t)$:
\begin{eqnarray}\label{eq:DeltaSc}
\langle \widehat{H}_S(\lambda_t) \overline{H}_S(\lambda_t) \rangle_t \dot{\Delta}_{Sc}= \mathcal{D}_t \langle \widehat{J}_c \rangle_t-{L}_{cc}(t)\Delta_{Sc},
\end{eqnarray}
where $\mathcal{D}_t=\big(\langle \widehat{H}_S(\lambda_t) \overline{H}_S(\lambda_t) \rangle_t +\langle \widehat{V}_c \overline{V}_c \rangle_t\big)\langle \widehat{V}_c \overline{V}_c \rangle_t^{-1}$.

\bibliography{qthermo}

\end{document}